\documentclass[journal,twoside]{IEEEtran}

\usepackage{todonotes}
\usepackage{siunitx}
\usepackage{etex}
\usepackage{amssymb}
\usepackage{graphicx}
\usepackage{multirow}
\usepackage{epstopdf}
\usepackage{algorithmic}
\usepackage{caption}
\usepackage{subcaption}
\usepackage{algorithm}
\usepackage{lineno}
\usepackage[cmex10]{amsmath}
\usepackage{float}
\usepackage{nicefrac}
\usepackage{mathtools}
\usepackage{pifont}
\usepackage{booktabs}
\usepackage{url}
\usepackage{tikz}
\usepackage{pgfplots}
\usepackage[english]{babel}
\usepackage[utf8]{inputenc}
\usepackage{amsthm}
\usepackage{gensymb}
\usepackage{xcolor}
\usepackage{todonotes}
\usepackage{textcomp}
\usepackage{adjustbox}
\usepackage{soul}
\usepackage{threeparttable}

\usetikzlibrary{positioning,shapes,shadows,arrows,backgrounds}
\DeclarePairedDelimiter\floor{\lfloor}{\rfloor}
\DeclarePairedDelimiter\ceil{\lceil}{\rceil}
\usetikzlibrary{decorations.markings, decorations.pathreplacing,patterns,fit,arrows}
\usetikzlibrary{automata, positioning,}
\usepgfplotslibrary{groupplots}

\newcommand{\iter}[1]{^{\left(#1\right)}}

\newcommand\yup{\ding{52}}
\newcommand\nope{\ding{56}}
\newcommand*\concat{\mathbin{\|}}
\newcommand\architect{\textsc{architect}}

\usepackage{cite}
\pgfplotsset{
	log x ticks with fixed point/.style={
		xticklabel={
			\pgfkeys{/pgf/fpu=true}
			\pgfmathparse{exp(\tick)}%
			\pgfmathprintnumber[fixed relative, precision=3]{\pgfmathresult}
			\pgfkeys{/pgf/fpu=false}
		}
	},
	log y ticks with fixed point/.style={
		yticklabel={
			\pgfkeys{/pgf/fpu=true}
			\pgfmathparse{exp(\tick)}%
			\pgfmathprintnumber[fixed relative, precision=3]{\pgfmathresult}
			\pgfkeys{/pgf/fpu=false}
		}
	}
}

\pgfplotsset{
	NR plot/.style={mark=o, mark options={color=blue}},
	JA  plot/.style={mark=x, mark options={color=red}},
	pmax plot/.style={mark=x, dashed, mark options={color=red, solid}},
	kmax plot/.style={mark=x, mark options={color=cyan}},
	pmaxNR plot/.style={mark=o, dashed, mark options={color=blue,solid}},
	kmaxNR plot/.style={mark=o, mark options={color=magenta}},
	fpt16 plot/.style={mark=o, mark options={color=green}},
	piso plot/.style={mark=x, mark options={color=blue}},
	ours plot/.style={mark=+, mark options={color=red}},
	para plot/.style={mark=o, mark options={color=blue}},
	paraandmsd plot/.style={mark=*, mark options={color=red}}
}

\newcommand{\func}[2]{#1{\left(#2\right)}}
\newcommand{\norm}[1]{\left\lVert#1\right\rVert}
\newcommand{\abs}[1]{\left|#1\right|}

\newcommand{\resettabcolsep}{\setlength{\tabcolsep}{5pt}}
\resettabcolsep

\begin{document}
	
	\title{\architect: Arbitrary-precision Hardware with Digit Elision for Efficient Iterative Compute}

	\author{
		He~Li,~\IEEEmembership{Student~Member,~IEEE},
		James~J.~Davis,~\IEEEmembership{Member,~IEEE}, 
		John~Wickerson,~\IEEEmembership{Senior~Member,~IEEE},
		and~George~A.~Constantinides,~\IEEEmembership{Senior~Member,~IEEE}%
		\thanks{
			The authors are with the Department of Electrical and Electronic Engineering, Imperial College London, London, SW7 2AZ, United Kingdom.
			E-mail: \texttt{\{h.li16, james.davis, j.wickerson, g.constantinides\}@imperial.ac.uk}.
		}
	}

	\markboth{IEEE Transactions on Very Large Scale Integration (VLSI) Systems}{Li \MakeLowercase{\emph{et al.}}: \architect: Arbitrary-precision Hardware with Digit Elision for Efficient Iterative Compute}
	
	\maketitle
	
	\begin{abstract}
		
		Many algorithms feature an iterative loop that converges to the result of interest.
		The numerical operations in such algorithms are generally implemented using finite-precision arithmetic, either fixed- or floating-point, most of which operate least-significant digit first.
		This results in a fundamental problem: if, after some time, the result has not converged, is this because we have not run the algorithm for enough iterations or because the arithmetic in some iterations was insufficiently precise?
		There is no easy way to answer this question, so users will often over-budget precision in the hope that the answer will always be to run for a few more iterations.
		We propose a fundamentally new approach: with the appropriate arithmetic able to generate results from most-significant digit first, we show that fixed compute-area hardware can be used to calculate an arbitrary number of algorithmic iterations to arbitrary precision, with both precision and approximant index increasing in lockstep.
		Consequently, datapaths constructed following our principles demonstrate efficiency over their traditional arithmetic equivalents where the latter's precisions are either under- or over-budgeted for the computation of a result to a particular accuracy.
		Use of most-significant digit-first arithmetic additionally allows us to declare certain digits to be stable at runtime, avoiding their recalculation in subsequent iterations and thereby increasing performance and decreasing memory footprints.
		Versus arbitrary-precision iterative solvers without the optimisations we detail herein, we achieve up-to 16$\times$ performance speedups and 1.9$\times$ memory savings for the evaluated benchmarks.

	\end{abstract}


	\section{Introduction}
	\label{sec:introduction}

		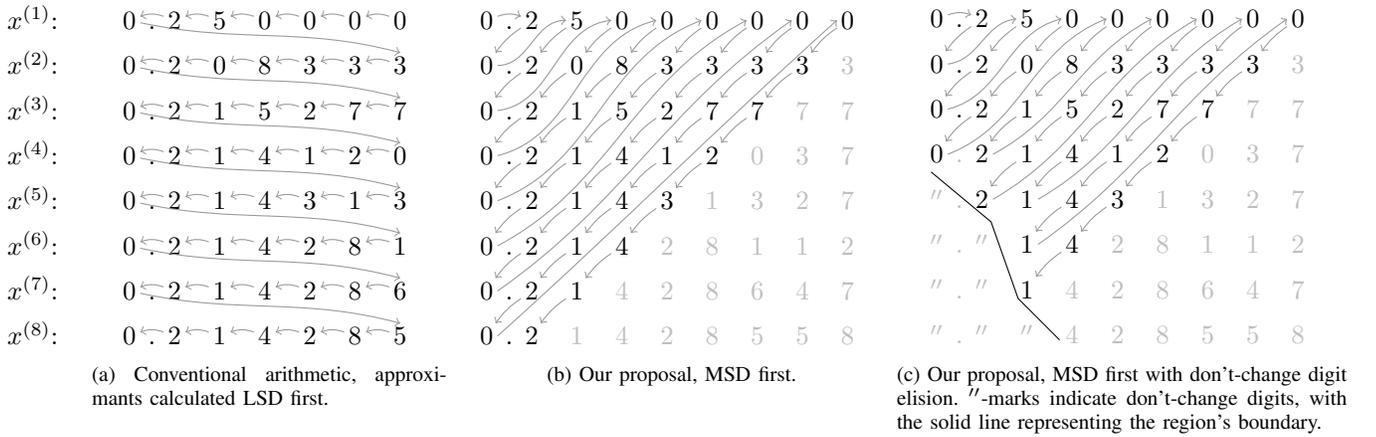
\begin{figure*}
			\centering
			\begin{subfigure}[t]{0.08\textwidth}
				\centering
				\begin{tikzpicture}[yscale=-1,inner sep=0.5mm,scale=0.6]
		
	\foreach \y in {1,...,8}
		\node at (-1.5,\y) {\vphantom{$1$}\smash{$x\iter{\y}{:}$}};
	
\end{tikzpicture}
			\end{subfigure}%
			\begin{subfigure}[t]{0.26\textwidth}
				\centering
				\begin{tikzpicture}[yscale=-1,inner sep=0.5mm,scale=0.6]

	\foreach \y in {0,...,7}
		\node(p\y) at (0.5,\y) {\vphantom{$1$}$.$};
			
	\node(x00) at (0,0) {$0$};
	\node(x10) at (1,0) {$2$};
	\node(x20) at (2,0) {$5$};
	\node(x30) at (3,0) {$0$};
	\node(x40) at (4,0) {$0$};
	\node(x50) at (5,0) {$0$};
	\node(x60) at (6,0) {$0$};
	
	\node(x01) at (0,1) {$0$};
	\node(x11) at (1,1) {$2$};
	\node(x21) at (2,1) {$0$};
	\node(x31) at (3,1) {$8$};
	\node(x41) at (4,1) {$3$};
	\node(x51) at (5,1) {$3$};
	\node(x61) at (6,1) {$3$};
	
	\node(x02) at (0,2) {$0$};
	\node(x12) at (1,2) {$2$};
	\node(x22) at (2,2) {$1$};
	\node(x32) at (3,2) {$5$};
	\node(x42) at (4,2) {$2$};
	\node(x52) at (5,2) {$7$};
	\node(x62) at (6,2) {$7$};
	
	\node(x03) at (0,3) {$0$};
	\node(x13) at (1,3) {$2$};
	\node(x23) at (2,3) {$1$};
	\node(x33) at (3,3) {$4$};
	\node(x43) at (4,3) {$1$};
	\node(x53) at (5,3) {$2$};
	\node(x63) at (6,3) {$0$};
	
	\node(x04) at (0,4) {$0$};
	\node(x14) at (1,4) {$2$};
	\node(x24) at (2,4) {$1$};
	\node(x34) at (3,4) {$4$};
	\node(x44) at (4,4) {$3$};
	\node(x54) at (5,4) {$1$};
	\node(x64) at (6,4) {$3$};
	
	\node(x05) at (0,5) {$0$};
	\node(x15) at (1,5) {$2$};
	\node(x25) at (2,5) {$1$};
	\node(x35) at (3,5) {$4$};
	\node(x45) at (4,5) {$2$};
	\node(x55) at (5,5) {$8$};
	\node(x65) at (6,5) {$1$};
	
	\node(x06) at (0,6) {$0$};
	\node(x16) at (1,6) {$2$};
	\node(x26) at (2,6) {$1$};
	\node(x36) at (3,6) {$4$};
	\node(x46) at (4,6) {$2$};
	\node(x56) at (5,6) {$8$};
	\node(x66) at (6,6) {$6$};
	
	\node(x07) at (0,7) {$0$};
	\node(x17) at (1,7) {$2$};
	\node(x27) at (2,7) {$1$};
	\node(x37) at (3,7) {$4$};
	\node(x47) at (4,7) {$2$};
	\node(x57) at (5,7) {$8$};
	\node(x67) at (6,7) {$5$};
		
	\foreach \y in {0,...,7} {
		\foreach[count=\xx] \x in {0,...,5} {
			\draw[->, black!40] (x\xx\y) to[bend left=20] (x\x\y);
		}
	}
	
	\foreach[count=\yy] \y in {0,...,6} {
		\newcommand\Ajohn{15}
		\draw[->, black!40] (x0\y) to[out=\Ajohn, in=\Ajohn-180] (x6\yy.north);
	}
	
\end{tikzpicture}
				\caption{Conventional arithmetic, approximants calculated LSD first.}
				\label{fig:digit_gen_trad}
			\end{subfigure}%
			\begin{subfigure}[t]{0.33\textwidth}
				\centering
				\begin{tikzpicture}[yscale=-1,inner sep=0.5mm,scale=0.6]
	
	\newcommand\gjohn{\color{black!25}}
			
	\foreach \y in {0,...,7}
		\node(p\y) at (0.5,\y) {\vphantom{$1$}$.$};

	\node(x00) at (0,0) {$0$};
	\node(x10) at (1,0) {$2$};
	\node(x20) at (2,0) {$5$};
	\node(x30) at (3,0) {$0$};
	\node(x40) at (4,0) {$0$};
	\node(x50) at (5,0) {$0$};
	\node(x60) at (6,0) {$0$};
	\node(x70) at (7,0) {$0$};
	\node(x80) at (8,0) {$0$};
	
	\node(x01) at (0,1) {$0$};
	\node(x11) at (1,1) {$2$};
	\node(x21) at (2,1) {$0$};
	\node(x31) at (3,1) {$8$};
	\node(x41) at (4,1) {$3$};
	\node(x51) at (5,1) {$3$};
	\node(x61) at (6,1) {$3$};
	\node(x71) at (7,1) {$3$};
	\node(x81) at (8,1) {\gjohn$3$};
	
	\node(x02) at (0,2) {$0$};
	\node(x12) at (1,2) {$2$};
	\node(x22) at (2,2) {$1$};
	\node(x32) at (3,2) {$5$};
	\node(x42) at (4,2) {$2$};
	\node(x52) at (5,2) {$7$};
	\node(x62) at (6,2) {$7$};
	\node(x72) at (7,2) {\gjohn$7$};
	\node(x82) at (8,2) {\gjohn$7$};
	
	\node(x03) at (0,3) {$0$};
	\node(x13) at (1,3) {$2$};
	\node(x23) at (2,3) {$1$};
	\node(x33) at (3,3) {$4$};
	\node(x43) at (4,3) {$1$};
	\node(x53) at (5,3) {$2$};
	\node(x63) at (6,3) {\gjohn$0$};
	\node(x73) at (7,3) {\gjohn$3$};
	\node(x83) at (8,3) {\gjohn$7$};
	
	\node(x04) at (0,4) {$0$};
	\node(x14) at (1,4) {$2$};
	\node(x24) at (2,4) {$1$};
	\node(x34) at (3,4) {$4$};
	\node(x44) at (4,4) {$3$};
	\node(x54) at (5,4) {\gjohn$1$};
	\node(x64) at (6,4) {\gjohn$3$};
	\node(x74) at (7,4) {\gjohn$2$};
	\node(x84) at (8,4) {\gjohn$7$};
	
	\node(x05) at (0,5) {$0$};
	\node(x15) at (1,5) {$2$};
	\node(x25) at (2,5) {$1$};
	\node(x35) at (3,5) {$4$};
	\node(x45) at (4,5) {\gjohn$2$};
	\node(x55) at (5,5) {\gjohn$8$};
	\node(x65) at (6,5) {\gjohn$1$};
	\node(x75) at (7,5) {\gjohn$1$};
	\node(x85) at (8,5) {\gjohn$2$};
	
	\node(x06) at (0,6) {$0$};
	\node(x16) at (1,6) {$2$};
	\node(x26) at (2,6) {$1$};
	\node(x36) at (3,6) {\gjohn$4$};
	\node(x46) at (4,6) {\gjohn$2$};
	\node(x56) at (5,6) {\gjohn$8$};
	\node(x66) at (6,6) {\gjohn$6$};
	\node(x76) at (7,6) {\gjohn$4$};
	\node(x86) at (8,6) {\gjohn$7$};
	
	\node(x07) at (0,7) {$0$};
	\node(x17) at (1,7) {$2$};
	\node(x27) at (2,7) {\gjohn$1$};
	\node(x37) at (3,7) {\gjohn$4$};
	\node(x47) at (4,7) {\gjohn$2$};
	\node(x57) at (5,7) {\gjohn$8$};
	\node(x67) at (6,7) {\gjohn$5$};
	\node(x77) at (7,7) {\gjohn$5$};
	\node(x87) at (8,7) {\gjohn$8$};
		
	\draw[->, black!40] (x00) to[bend right=20] (x10);
		
	\foreach \x [
		evaluate=\x as \ang using int((\x-1)*7),
		evaluate=\x as \y using int(\x+1)
	] in {1,...,7} {
		\draw[->, black!40] (x0\x.east) to[out=-\ang, in=-180-\ang] (x\y0.west);
	}
	
	\foreach \d in {1,...,7} {
		\foreach \x [
			evaluate=\x as \xnew using int(\x-1),
			evaluate=\x as \y using int(\d-\x),
			evaluate=\x as \ynew using int(\d-\x+1)
		] in {\d,...,1} {
			\draw[->, black!40] (x\x\y) to[bend left=10] (x\xnew\ynew);
		}
	}
	\foreach \x [
		evaluate=\x as \xnew using int(\x-1),
		evaluate=\x as \y using int(8-\x),
		evaluate=\x as \ynew using int(8-\x+1)
	] in {8,...,2} {
		\draw[->, black!40] (x\x\y) to[bend left=10] (x\xnew\ynew);
	}
	
\end{tikzpicture}
				\caption{Our proposal, MSD first.}
				\label{fig:digit_gen_arch}
			\end{subfigure}%
			\begin{subfigure}[t]{0.33\textwidth}
				\centering
				\begin{tikzpicture}[yscale=-1,inner sep=0.5mm,scale=0.6]
	
	\newcommand\gjohn{\color{black!25}}

	\foreach \y in {0,...,2}
		\node(p\y) at (0.5,\y) {\vphantom{$1$}$.$};
	\foreach \y in {3,...,7}
		\node(p\y) at (0.5,\y) {\vphantom{$1$}\gjohn$.$};

	\node(x01) at (0,0) {$0$};
	\node(x11) at (1,0) {$2$};
	\node(x21) at (2,0) {$5$};
	\node(x31) at (3,0) {$0$};
	\node(x41) at (4,0) {$0$};
	\node(x51) at (5,0) {$0$};
	\node(x61) at (6,0) {$0$};
	\node(x71) at (7,0) {$0$};
	\node(x81) at (8,0) {$0$};

	\node(x02) at (0,1) {$0$};
	\node(x12) at (1,1) {$2$};
	\node(x22) at (2,1) {$0$};
	\node(x32) at (3,1) {$8$};
	\node(x42) at (4,1) {$3$};
	\node(x52) at (5,1) {$3$};
	\node(x62) at (6,1) {$3$};
	\node(x72) at (7,1) {$3$};
	\node(x82) at (8,1) {\gjohn$3$};

	\node(x03) at (0,2) {$0$};
	\node(x13) at (1,2) {$2$};
	\node(x23) at (2,2) {$1$};
	\node(x33) at (3,2) {$5$};
	\node(x43) at (4,2) {$2$};
	\node(x53) at (5,2) {$7$};
	\node(x63) at (6,2) {$7$};
	\node(x73) at (7,2) {\gjohn$7$};
	\node(x83) at (8,2) {\gjohn$7$};

	\node(x04) at (0,3) {$0$};
	\node(x14) at (1,3) {$2$};
	\node(x24) at (2,3) {$1$};
	\node(x34) at (3,3) {$4$};
	\node(x44) at (4,3) {$1$};
	\node(x54) at (5,3) {$2$};
	\node(x64) at (6,3) {\gjohn$0$};
	\node(x74) at (7,3) {\gjohn$3$};
	\node(x84) at (8,3) {\gjohn$7$};

	\node(x05) at (0,4) {\gjohn$''$};
	\node(x15) at (1,4) {$2$};
	\node(x25) at (2,4) {$1$};
	\node(x35) at (3,4) {$4$};
	\node(x45) at (4,4) {$3$};
	\node(x55) at (5,4) {\gjohn$1$};
	\node(x65) at (6,4) {\gjohn$3$};
	\node(x75) at (7,4) {\gjohn$2$};
	\node(x85) at (8,4) {\gjohn$7$};

	\node(x06) at (0,5) {\gjohn$''$};
	\node(x16) at (1,5) {\gjohn$''$};
	\node(x26) at (2,5) {$1$};
	\node(x36) at (3,5) {$4$};
	\node(x46) at (4,5) {\gjohn$2$};
	\node(x56) at (5,5) {\gjohn$8$};
	\node(x66) at (6,5) {\gjohn$1$};
	\node(x76) at (7,5) {\gjohn$1$};
	\node(x86) at (8,5) {\gjohn$2$};

	\node(x07) at (0,6) {\gjohn$''$};
	\node(x17) at (1,6) {\gjohn$''$};
	\node(x27) at (2,6) {$1$};
	\node(x37) at (3,6) {\gjohn$4$};
	\node(x47) at (4,6) {\gjohn$2$};
	\node(x57) at (5,6) {\gjohn$8$};
	\node(x67) at (6,6) {\gjohn$6$};
	\node(x77) at (7,6) {\gjohn$4$};
	\node(x87) at (8,6) {\gjohn$7$};

	\node(x08) at (0,7) {\gjohn$''$};
	\node(x18) at (1,7) {\gjohn$''$};
	\node(x28) at (2,7) {\gjohn$''$};
	\node(x38) at (3,7) {\gjohn$4$};
	\node(x48) at (4,7) {\gjohn$2$};
	\node(x58) at (5,7) {\gjohn$8$};
	\node(x68) at (6,7) {\gjohn$5$};
	\node(x78) at (7,7) {\gjohn$5$};
	\node(x88) at (8,7) {\gjohn$8$};

	\draw[shorten <=-1mm, shorten >=-1mm] (0,3.5) to (1.2,4.5) to (1.8,6.2) to (2.6,7);

	\draw[->, black!40] (x00) to[bend right=20] (x10);
	
	\foreach \x [
	evaluate=\x as \ang using int((\x-1)*7),
	evaluate=\x as \y using int(\x)
	] in {2,...,4} {
		\draw[->, black!40] (x0\x.east) to[out=-\ang, in=-180-\ang] (x\y0.west);
	}
	\foreach \x [
	evaluate=\x as \ang using int((\x-1)*7),
	evaluate=\x as \y using int(\x),
	evaluate=\x as \xmsd using int(\x-1)
	] in {5,...,5} {
		\draw[->, black!40] (x1\xmsd.east) to[out=-\ang, in=-180-\ang] (x\y0.west);
	}
	\foreach \x [
	evaluate=\x as \ang using int((\x-1)*6),
	evaluate=\x as \y using int(\x),
	evaluate=\x as \xmsd using int(\x-1)
	] in {6,...,6} {
		\draw[->, black!40] (x1\xmsd.east) to[out=-\ang, in=-180-\ang] (x\y0.west);
	}
	\foreach \x [
	evaluate=\x as \ang using int((\x-1)*5),
	evaluate=\x as \y using int(\x),
	evaluate=\x as \xmsd using int(\x-2)
	] in {7,...,7} {
		\draw[->, black!40] (x2\xmsd.east) to[out=-\ang, in=-180-\ang] (x\y0.west);
	}
	\foreach \x [
	evaluate=\x as \ang using int((\x-1)*5),
	evaluate=\x as \y using int(\x),
	evaluate=\x as \xmsd using int(\x-2)
	] in {8,...,8} {
		\draw[->, black!40] (x2\xmsd.east) to[out=-\ang, in=-180-\ang] (x\y0.west);
	}

	\foreach \d in {1,...,3} {
		\foreach \x [
		evaluate=\x as \xnew using int(\x-1),
		evaluate=\x as \y using int(\d+1-\x),
		evaluate=\x as \ynew using int(\d+1-\x+1)
		] in {\d,...,1} {
			\draw[->, black!40] (x\x\y) to[bend left=10] (x\xnew\ynew);
		}
	}
	\foreach \x [
	evaluate=\x as \xnew using int(\x-1),
	evaluate=\x as \y using int(5-\x),
	evaluate=\x as \ynew using int(5-\x+1)
	] in {4,...,2} {
		\draw[->, black!40] (x\x\y) to[bend left=10] (x\xnew\ynew);
	}
	\foreach \x [
	evaluate=\x as \xnew using int(\x-1),
	evaluate=\x as \y using int(6-\x),
	evaluate=\x as \ynew using int(6-\x+1)
	] in {5,...,2} {
		\draw[->, black!40] (x\x\y) to[bend left=10] (x\xnew\ynew);
	}
	\foreach \x [
	evaluate=\x as \xnew using int(\x-1),	
	evaluate=\x as \y using int(7-\x),
	evaluate=\x as \ynew using int(7-\x+1)
	] in {6,...,3} {
		\draw[->, black!40] (x\x\y) to[bend left=10] (x\xnew\ynew);
	}
	\foreach \x [
	evaluate=\x as \xnew using int(\x-1),
	evaluate=\x as \y using int(8-\x),
	evaluate=\x as \ynew using int(8-\x+1)
	] in {7,...,3} {
		\draw[->, black!40] (x\x\y) to[bend left=10] (x\xnew\ynew);
	}
	\foreach \x [
	evaluate=\x as \xnew using int(\x-1),
	evaluate=\x as \y using int(9-\x),
	evaluate=\x as \ynew using int(9-\x+1)
	] in {8,...,3} {
		\draw[->, black!40] (x\x\y) to[bend left=10] (x\xnew\ynew);
	}
	
\end{tikzpicture}
				\caption{
					Our proposal, MSD first with don't-change digit elision.
					$''$-marks indicate don't-change digits, with the solid line representing the region's boundary.
				}
				\label{fig:digit_gen_prop}
			\end{subfigure}
			\caption{
				Digit-calculating strategies for the solution of $x\iter{k+1} = \nicefrac{1}{4} - \nicefrac{1}{6}\cdot x\iter{k}$ starting from $x\iter{0} = 0$.
				Arrows show the order of digit generation.
			}
			\label{fig:digit_gen}
		\end{figure*}
		
		\IEEEPARstart{I}{n} numerical analysis, an algorithm executing on the real numbers, $\mathbb{R}$, is often expressed as a conceptually infinite iterative process that converges to a result.
		This is illustrated in a general form by the equation
		\begin{equation*}
		\boldsymbol{x}\iter{k + 1} = f{\left(\boldsymbol{x}\iter{k}\right)},
		\end{equation*}
		in which the computable real function $f \in \left(\mathbb{R}^N \to \mathbb{R}^N\right)$ is repeatedly applied to an initial approximation $\boldsymbol{x}\iter{0} \in \mathbb{R}^N$.
		The true result, $\boldsymbol{x}^*$, is obtained as $k$ approaches infinity, \emph{i.e.}
		\begin{equation*}
			\boldsymbol{x}^* = \lim_{k \to \infty} \Pi{\left(\boldsymbol{x}\iter{k}\right)},
		\end{equation*}
		where the operator $\Pi$ denotes projection of the variables of interest since the result may be of lower dimensionality than $N$.
		Examples of this template include classical iterative methods such as Jacobi and successive over-relaxation, as well as others including gradient descent methods, the key algorithms in deep learning~\cite{lecun2015deep}.
		
		In practice, these calculations are often implemented using finite-precision approximations such as that shown in Algorithm~\ref{alg:fpi}, wherein $\mathbb{FP}_P$ denotes some finite-precision datatype, $P$ is a measure of its precision (usually word length), $\hat{f}$ is a finite-precision approximation of $f$ and $\eta$ is an accuracy bound.
		The problem with this implementation lies in the coupling of $P$ and iteration limit $K$.
		Generally, this algorithm will \emph{not} be able to ensure that its assertion passes, and when it fails we are left with no knowledge as to whether $K$ should be increased or if all computations need to be thrown away and the algorithm restarted with a higher $P$ instead.
		
		\renewcommand{\algorithmicensure}{\textbf{Assert:}}
		
		\begin{algorithm}[t]
			\begin{algorithmic}[1]
				\REQUIRE $\hat{\boldsymbol{x}}\iter{0} \in \mathbb{FP}_P^N$, $\hat{f} \in \left(\mathbb{FP}_P^N \to \mathbb{FP}_P^N\right)$
				\FOR{$k = 0$ \TO $K - 1$}
					\STATE $\hat{\boldsymbol{x}}\iter{k + 1} \gets \hat{f}{\left(\hat{\boldsymbol{x}}\iter{k}\right)}$
				\ENDFOR
				\ENSURE $\left\|\Pi{\left(\hat{\boldsymbol{x}}\iter{K}\right)} - \boldsymbol{x}^*\right\| < \eta$
			\end{algorithmic}
			\caption{Generic finite-precision iterative algorithm.}
			\label{alg:fpi}
		\end{algorithm}		
		\IEEEpubidadjcol
		
		As a simple demonstration of this problem, suppose we wish to compute the toy iteration
		\begin{equation*}
			x\iter{k + 1} = \nicefrac{1}{4} - \nicefrac{1}{6}\cdot x\iter{k}
		\end{equation*}	
		starting from zero.
		
		When performing this arithmetic using a standard approach in either software or hardware, we must choose a single, fixed precision for our calculations before beginning to iterate.
		Fig.~\ref{fig:digit_gen_trad} shows the order in which digits are calculated when the precision is fixed to six decimal places: approximant-by-approximant, least-significant digit (LSD) first.
		Choosing the right precision \emph{a priori} is difficult, particularly with respect to hardware implementation.
		If it is too high, the circuit may be unnecessarily slow and power-hungry, while, if it is too low, the criterion for convergence may never be reached.
		
		Our proposal, illustrated in Fig.~\ref{fig:digit_gen_arch}, avoids the need to answer the aforementioned question entirely.
		The digits are calculated in a zig-zag pattern, sweeping through approximants and decimal places simultaneously.
		The longer we compute, the more accurate our result will be; the computation can terminate whenever the result is accurate enough.
		This avoids the need to fix the precision beforehand, but requires the ability to calculate from most-significant digit (MSD) first: a facility provided through the use of \emph{online arithmetic}~\cite{OAbook}.
		While general-purpose processors featuring traditional, LSD-first arithmetic units exhibit inefficiency for the realisation of online arithmetic, field-programmable gate arrays (FPGAs) represent excellent platforms for the implementation of such MSD-first operations~\cite{KanFPT14,fpt16,ARCHITECT}.

		As originally formulated, this method is somewhat inefficient since the triangular shape traced out results in the computation of more digits than is actually needed.
		In the bottom-left corner lie high-significance digits of later approximants; these generally become stable over time, so we call them \emph{don't-change} digits. 
		By detecting the presence and avoiding the recomputation of these digits, we arrive at a digit pattern such as that shown in Fig.~\ref{fig:digit_gen_prop}.
		This increases efficiency while having no bearing on the chosen iterative method's ability to reach a result of any accuracy.
		
		The proposed architecture, coined \architect{} (for \textbf{Ar}bitrary-precision \textbf{C}onstant-\textbf{h}ardware \textbf{Ite}rative \textbf{C}ompu\textbf{t}e), is the first to allow the runtime adaption of both precision and iteration count for iterative algorithms implemented in hardware.
        We make the following novel contributions:	
		\begin{itemize}
			\item The first fixed-compute-resource hardware for iterative calculation capable of producing arbitrary-precision results after arbitrary numbers of iterations.
			\item An optimised mechanism for digit-vector storage based on a Cantor pairing function to facilitate simultaneously increasing precision and iteration count.
			\item Theoretical analysis of MSD stability within any online arithmetic-implemented iterative method, enabling the runtime elision of don't-change digits to obtain performance speedups and increase memory efficiency.
			\item Exemplary hardware implementations of our proposals for the computation of both linear (Jacobi method) and nonlinear (Newton) iterations.
			\item Qualitative and quantitative performance and scalability comparisons against traditional and state-of-the-art online arithmetic FPGA implementations.
		\end{itemize}
		
		An earlier version of this work appeared in the proceedings of the 16\textsuperscript{th} International Conference on Field-programmable Technology (FPT)~\cite{ARCHITECT}.
		This article combines that paper's material with the don't-change digit proposal taken from our 24\textsuperscript{th} IEEE Symposium on Computer Arithmetic (ARITH) publication~\cite{lidigit}, extending both.
		In particular:
		\begin{itemize}
			\item We add an arbitrary-precision divider to our available operators, enabling the construction of datapaths for the calculation of irrational results with Newton's method.
			\item Changes to our digit elision technique, originally designed for linear-convergence algorithms, are made to suit the Newton method's quadratic convergence.
			\item To complement the new digit elision strategy, we propose an enhanced memory-addressing scheme, leading to greater performance and higher achievable result accuracy for a given memory budget.
			\item Finally, we exploit digit-parallel online addition to decrease datapath latency.
		\end{itemize}
		These optimisations allow us to obtain significant increases in throughput and memory efficiency over previous designs.
		
		The implementations presented and evaluated in this article are fixed-point.
		\architect{}'s principles are, however, generic, and could be employed for the construction of floating-point operators supporting arbitrary-precision mantissas.

	\section{Background}
	\label{sec:background}
			
		In scientific computing, machine learning, optimisation and many other numerical application areas, methods of iterative calculation are particularly popular and interest in their acceleration with FPGAs is growing~\cite{iter_acc}.
		Recent studies have demonstrated that FPGAs are promising platforms for the acceleration of the Jacobi~\cite{TC08}, Newton's~\cite{liu2016fpga}, conjugate gradient~\cite{roldao2009} and MINRES methods~\cite{boland2008MINRES}.
		However, implementations relying on traditional arithmetic---whether digit-serial or -parallel---enforce compile-time determination of precision; for digit-parallel designs this affects their area and input/output bandwidth requirements, while for digit-serial it is one of the factors affecting algorithm runtime.
		Runtime tuning of precision in iterative calculations was enabled through the use of online arithmetic in recent work~\cite{fpt16}, however unrolling was necessary in order to implement the algorithm's loop; area therefore scaled with the desired number of iterations.
		As shown in Table~\ref{tab:arith_comp}, \architect{} stands apart from these alternatives by enabling the runtime selection of both factors affecting result accuracy while keeping compute area constant.
		
		\begin{table}
			\caption{Comparison of iterative arithmetic paradigms.}
			\centering
			\begin{tabular}{ccccc}
				\toprule
												&	\multicolumn{2}{c}{Area scales with}	&	\multicolumn{2}{c}{Runtime scales with}	\\
													\cmidrule(lr){2-3}							\cmidrule(lr){4-5}
												&	Prec.	&	Iter. limit					&	Prec.			&	Iter. limit			\\ 
				\midrule
				LSD-first, parallel				&	\yup	&	\nope						&	\nope			&	\yup~unbounded		\\
				LSD-first, serial				&	\nope	&	\nope						&	\yup~bounded	&	\yup~unbounded		\\
				Zhao \emph{et al.}~\cite{fpt16}	&	\nope	&	\yup						&	\yup~unbounded	&	\nope				\\
				\architect{}					&	\nope	&	\nope						&	\yup~unbounded	&	\yup~unbounded		\\ 
				\bottomrule
			\end{tabular}
			\label{tab:arith_comp}
		\end{table}

		\subsection{Arbitrary-precision Computing}
		\label{subsec:arb_prec}
						
			Applications requiring very high precisions have become increasingly popular in recent years~\cite{DT11}.
			For example, today, hundreds of digits of precision are required in atomic system simulations and electromagnetic scattering theory calculations, while Ising integrals and elliptic function evaluation need thousands of digits~\cite{highprec12}.
			In experimental mathematics, Poisson equation computations frequently require results to tens or hundreds of thousands of digits of precision~\cite{threadarb}.
			Standard numeric datatypes, such as double- or even quadruple-precision floating-point, are therefore no longer sufficient in an increasing number of scenarios.		
			
			Many software libraries have been developed for arbitrary-precision arithmetic~\cite{serpette1989bignum,mpfr,johansson2017arb}.
			The \emph{de facto} standard is MPFR, which guarantees correct rounding to any requested number of bits, selected before each operation is executed.
			Interest in the hardware acceleration of high-precision operations, in particular those within iterative algorithms, is growing~\cite{iter_acc}.
			FPGAs provide flexibility not available on other platforms, allowing for the implementation of bespoke designs with many precision and performance tradeoffs.
			Libraries including FloPoCo~\cite{FloPoCo} and VFLOAT~\cite{TRETS16}, alongside proprietary vendor tools, facilitate the creation of custom-precision arithmetic units.
			These provide designers with many options to suit particular frequency, latency and resource requirements.
			Sun \emph{et al.} proposed an FPGA-based mixed-precision linear solver: as many operations as possible are performed in low precision before switching to a slower, higher-precision mode for the later iterations~\cite{TC08}.
			A dual-mode (double- and quadruple-precision) architecture based on Taylor series expansion has also been implemented~\cite{jaiswal2017dualmode}.
			Zhao \emph{et al.}'s work enables arbitrary-precision computation but, as mentioned earlier, necessitates compile-time determination of iteration count~\cite{fpt16}.
			
			With the exception of Zhao \textit{et al.}'s architecture~\cite{fpt16}, each of the aforementioned proposals requires precision---or precisions---to be determined \emph{a priori}.
			In many cases, this is not a trivial task; making the wrong choice often means having to throw the calculations already done away and starting from scratch with higher precision, wasting both time and energy in doing so.
			In our work, we are particularly interested in hardware architectures which allow precision to be increased over time without having to restart computation or modify the circuitry.
			Table~\ref{tab:comp} presents a side-by-side comparison of these techniques and their features with \architect{}, the only entry supporting the determination of result precision and iteration count \emph{after each calculation has commenced.}
			
			\begin{table}
				\caption{Comparison of arbitrary-precision techniques.}
				\centering
				\begin{tabular}{ccccc}
					\toprule
																		&	\multirow{2}{*}{Level}	&	\multirow{2}{*}{\shortstack{Prec. set\\per calc.}}		&	\multirow{2}{*}{\shortstack{Iter. limit set\\per calc.}}	\\
																		&							&															&			\\
					\midrule
					MPFR~\cite{mpfr}									&	Software				&	Before													&	During	\\
					FloPoCo~\cite{FloPoCo}, \emph{etc.}					&	Hardware				&	Before													&	During	\\
					Mixed precision~\cite{TC08, jaiswal2017dualmode}	&	Hardware				&	Before													&	During	\\
					Zhao \emph{et al.}~\cite{fpt16}						&	Hardware				&	During													&	Before	\\
					\architect{}										&	Hardware				&	During													&	During	\\
					\bottomrule
				\end{tabular}
				\label{tab:comp}
			\end{table}
		
	    \subsection{Online Arithmetic}
		\label{subsec:online}
			
			Achieving arbitrary-precision computation with fixed hardware requires MSD-first input consumption and output generation.
			A suitable proposal for this, widely discussed in the literature, is online arithmetic~\cite{OAbook}.
			By employing redundancy in their number representation, allowing less-significant digits to correct errors introduced in those of higher significance, all online operators are able to function in MSD-first fashion.
			Online operators are classically serial, however efficient digit-parallel (unrolled) implementations targetting FPGAs have been developed as well~\cite{KanFPT14}.
			We make use of both digit-serial and -parallel online operators in this work, employing the \emph{de facto} standard radix-2 signed-digit number representation, wherein the $i^\textnormal{th}$ digit of a number $x$, $x_i$, lies in $\left\{-1, 0, 1\right\}$.
			In hardware, each $x_i$ corresponds to a pair of bits, $x_i^{+}$ and $x_i^{-}$, selected such that $x_i = x_i^{+} - x_i^{-}$.
			Data can be efficiently converted between non-redundant and redundant forms using well known on-the-fly conversion techniques~\cite{OAbook}.	
			
			Of particular significance to the material presented in this article is the concept of \emph{online delay}.
			When performing an online operation, the digits of its result are generated at the same rate as its input digits are consumed, but the result is delayed by a fixed number of digits, denoted $\delta$.
			That is, the first (\emph{i.e.} most-significant) $q$ digits of an operator's result are wholly determined by the first $q + \delta$ digits within each of its operands~\cite{OAbook}.
			The value of $\delta$ is operator-specific, but is typically a small integer.
			When chaining operators to form a datapath, as we do, the total online delay is the highest cumulative delay through the complete circuit~\cite{fpt16}.
			
			\subsubsection{Addition}
			\label{subsubsec:oa}
			
\begin{figure}
					\centering
					\tikzset{
	fad/.style={draw, minimum width=1.6cm, line width=0.8pt, minimum height=6mm},					
	del/.style={draw, inner sep=0.8mm, line width=0.8pt, minimum width=3mm, minimum height=3mm},	
	lbl/.style={inner sep=1.2pt, anchor=south}													
}

\newcommand\delaymarker[1]{
	\node[anchor=west, inner sep=0] at (#1.west) {\hspace*{-0.5pt}\tiny $>$};
}

\resizebox{\columnwidth}{!}{%
	\begin{tikzpicture}[yscale=-1]
	
	\small
	\node(xinp) at (0.0,0.1) {$x_\textnormal{in}^+$};
	\node(xinm) at (0.5,0.1) {$x_\textnormal{in}^-$};
	\node(yinp) at (1.0,0.1) {$y_\textnormal{in}^+$};
	\node(yinm) at (1.5,0.1) {$y_\textnormal{in}^-$};
	\node(zoutp) at (0.25,3.7) {$z_\textnormal{out}^+$};
	\node(zoutm) at (0.75,3.7) {$z_\textnormal{out}^-$};
	
	\node[fad] (fad1) at (0.5,1) {\raisebox{1.5mm}{Full adder}};
	\node[lbl] at (fad1.south -| zoutp) {$s$};
	\node[lbl] at (fad1.south -| zoutm) {$c$};
	\node[fad] (fad2) at (0.5,2.5) {\raisebox{1.5mm}{Full adder}};
	\node[lbl] at (fad2.south -| zoutp) {$s$};
	\node[lbl] at (fad2.south -| zoutm) {$c$};
	\node[del] (del1) at (0.75,1.6) {}; \delaymarker{del1}
	\node[del] (del2) at (1.5,1.6) {};  \delaymarker{del2}
	\node[del] (del3) at (0.75,3.1) {}; \delaymarker{del3}
	
	\draw     (xinp) to (fad1.north -| xinp);
	\draw[-o] (xinm) to (fad1.north -| xinm);
	\draw     (yinp) to (fad1.north -| yinp);
	\draw     (yinm) to (del2);
	\draw     (del1) to (fad1.south -| del1);
	\draw     (fad2.north -| xinp) to +(0,-0.2) to ([yshift=1mm]del1.south) to (del1.south);
	\draw[o-] (fad2.north -| xinm) to +(0,-0.2) to ([yshift=1mm]del2.south) to (del2.south);
	\draw     (fad2.north -| yinp) to +(0,-0.2) to ([yshift=1mm]del1.south -| zoutp) to (fad1.south -| zoutp);
	\draw[-o] (zoutp) to (fad2.south -| 0.25,4.2);
	\draw     (del3) to (fad2.south -| del3);
	\draw     (del3) to (zoutm);
	
	\begin{scope}[xshift=2.5cm]
	\node(x0p) at (0.0,0.1) {$x_0^+$};
	\node(x0m) at (0.5,0.1) {$x_0^-$};
	\node(y0p) at (1.0,0.1) {$y_0^+$};
	\node(y0m) at (1.5,0.1) {$y_0^-$};
	\node(x1p) at (2.0,0.1) {$x_1^+$};
	\node(x1m) at (2.5,0.1) {$x_1^-$};
	\node(y1p) at (3.0,0.1) {$y_1^+$};
	\node(y1m) at (3.5,0.1) {$y_1^-$};
	\node      at (4.0,0.1) {$\dots$};
	\node(x2p) at (4.5,0.1) {$x_{\!\scalebox{0.5}{$P{-}1$}}^+$};
	\node(x2m) at (5.0,0.1) {$x_{\!\scalebox{0.5}{$P{-}1$}}^-$};
	\node(y2p) at (5.5,0.1) {$y_{\!\scalebox{0.5}{$P{-}1$}}^+$};
	\node(y2m) at (6.0,0.1) {$y_{\!\scalebox{0.5}{$P{-}1$}}^-$};
	\node(cip) at (6.5,0.1) {$c_\textnormal{in}^+$};
	\node(cim) at (7.0,0.1) {$c_\textnormal{in}^-$};
	\node(cop) at (0.25,3.7) {$c_\textnormal{out}^+$};
	\node(com) at (0.75,3.7) {$c_\textnormal{out}^-$};
	\node(z0p) at (2.00,3.7) {$z_0^+$};
	\node(z0m) at (2.50,3.7) {$z_0^-$};
	\node(z1p) at (4.00,3.7) {$z_1^+$};
	\node      at (4.50,3.7) {$\dots$};
	\node(z1m) at (5.00,3.7) {$z_{\scalebox{0.5}{$P{-}2$}}^-$};
	\node(z2p) at (6.50,3.7) {$z_{\scalebox{0.5}{$P{-}1$}}^+$};
	\node(z2m) at (7.00,3.7) {$z_{\scalebox{0.5}{$P{-}1$}}^-$};
	
	\node[fad] (fad3) at (0.5,1) {\raisebox{1.5mm}{Full adder}};
	\node[lbl] at (fad3.south -| 0.25,0) {$s$};
	\node[lbl] at (fad3.south -| 0.75,0) {$c$};
	\node[fad] (fad5) at (2.5,1) {\raisebox{1.5mm}{Full adder}};
	\node[lbl] at (fad5.south -| 2.25,0) {$s$};
	\node[lbl] at (fad5.south -| 2.75,0) {$c$};
	\node[fad] (fad7) at (5.0,1) {\raisebox{1.5mm}{Full adder}};
	\node[lbl] at (fad7.south -| 4.75,0) {$s$};
	\node[lbl] at (fad7.south -| 5.25,0) {$c$};
	\node[fad] (fad4) at (1.5,2.5) {\raisebox{1.5mm}{Full adder}};
	\node[lbl] at (fad4.south -| 1.25,0) {$s$};
	\node[lbl] at (fad4.south -| 1.75,0) {$c$};
	\node[fad] (Fad4) at (3.5,2.5) {\raisebox{1.5mm}{Full adder}};
	\node[lbl] at (Fad4.south -| 3.25,0) {$s$};
	\node[lbl] at (Fad4.south -| 3.75,0) {$c$};
	\node[fad] (fad8) at (6.0,2.5) {\raisebox{1.5mm}{Full adder}};
	\node[lbl] at (fad8.south -| 5.75,0) {$s$};
	\node[lbl] at (fad8.south -| 6.25,0) {$c$};
	
	\draw     (x0p) to (fad3.north -| x0p);
	\draw[-o] (x0m) to (fad3.north -| x0m);
	\draw     (y0p) to (fad3.north -| y0p);
	\draw[-o] (y0m) to (fad4.north -| y0m);
	\draw     (x1p) to (fad5.north -| x1p);
	\draw[-o] (x1m) to (fad5.north -| x1m);
	\draw     (y1p) to (fad5.north -| y1p);
	\draw[-o] (y1m) to (Fad4.north -| y1m);
	\draw     (x2p) to (fad7.north -| x2p);
	\draw[-o] (x2m) to (fad7.north -| x2m);
	\draw     (y2p) to (fad7.north -| y2p);
	\draw[-o] (y2m) to (fad8.north -| y2m);
	\draw     (cop) to (fad3.south -| cop);
	\draw[-o] (com) |- (1.25,3.2) -| (fad4.south -| 1.25,0);
	\draw     (z0p) |- (1.75,3.2) -| (fad4.south -| 1.75,0);
	\draw[-o] (z0m) |- (3.25,3.2) -| (Fad4.south -| 3.25,0);
	\draw     (z1p) |- (3.75,3.2) -| (Fad4.south -| 3.75,0);
	\draw[-o] (z1m) |- (5.75,3.2) -| (fad8.south -| 5.75,0);
	\draw     (z2p) |- (6.25,3.2) -| (fad8.south -| 6.25,0);
	\draw     (cip) to (fad8.north -| cip);
	\draw     (cim) to (z2m);
	\draw     (fad3.south -| 0.75,0) |- (0.75,1.8) -| (fad4.north -| y0p);
	\draw     (fad5.south -| 2.25,0) |- (2.25,1.8) -| (fad4.north -| x1p);
	\draw     (fad5.south -| 2.75,0) |- (2.75,1.8) -| (Fad4.north -| y1p);
	\draw     (fad7.south -| 4.75,0) |- (4.55,1.8);
	\draw[dotted] (4.55,1.8) to (4.2,1.8);
	\draw     (4.2,1.8) -| (Fad4.north -| 4.0,0);
	\draw     (fad7.south -| 5.25,0) |- (5.5,1.8) -| (fad8.north -| y2p);
	\end{scope}
	
	\end{tikzpicture}%
}
					\caption{
						Radix-2 online adders~\cite{OAbook}.
						Left: Serial.
						Right: Parallel.
					}
					\label{fig:oa}
				\end{figure}
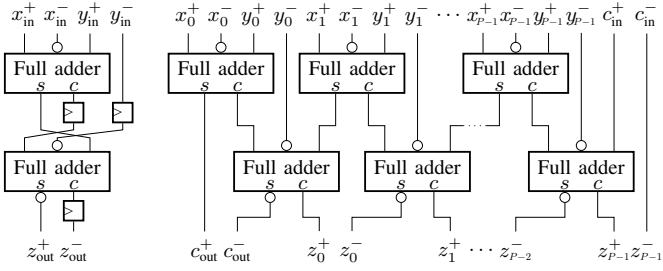

				A classical online adder makes use of full adders and registers to add digits of inputs $x$ and $y$ presented serially as $x_\textnormal{in}$ and $y_\textnormal{in}$, as shown in Fig.~\ref{fig:oa}~(left), from most to least significant~\cite{OAbook}.
				Digits of $z$ start to appear at serial output $z_\textnormal{out}$ after two clock cycles, hence $\delta_+ = 2$.
				Duplication of such a serial adder $P$ times and removal of its registers leads to the creation of a $P$-digit parallel online adder devoid of online delay, as shown in Fig.~\ref{fig:oa}~(right).
				Crucially, while carry digits are presented at the least-significant end of the adder and generated at the most, there is no carry chain; independent of its word length, the critical path lies across two full adders.
				This demonstrates the adder's suitability for the construction of more complex online operators and indicates that its maximum frequency is independent of precision.	
				
			\subsubsection{Multiplication}
			\label{subsubsec:om}
				
				Algorithm~\ref{alg:classic_om} illustrates classical radix-2 online multiplication~\cite{OAbook}: a process that operates in serial-in, serial-out fashion.
				So-called digit vectors $\boldsymbol{x}$ and $\boldsymbol{y}$ are assembled from digits of multiplicand $x$ and multiplier $y$ over time from the most significant first; $\concat$ represents concatenation performed such that
				\begin{equation*}
					\boldsymbol{x} = \sum_{i = 0}^{j}{x_i 2^{-i - 1}},
					\quad
					\boldsymbol{y} = \sum_{i = 0}^{j}{y_i 2^{-i - 1}}
				\end{equation*}
				during cycle $j$.
				Digit-selection function $\textnormal{sel}_\times$ serves to determine the digits of product $z$.
				This is defined to be
				\begin{equation*}
					\textnormal{sel}_\times{\left(\boldsymbol{v}\right)} =
					\begin{cases}
						1	&	\textnormal{if}~\boldsymbol{v} \geq \nicefrac{1}{2}						\\
						0	&	\textnormal{if}~-\nicefrac{1}{2} \leq \boldsymbol{v} < \nicefrac{1}{2}	\\
						-1	&	\textnormal{otherwise}.
					\end{cases}
				\end{equation*}
				$z_j$ is produced at cycle $j + 3$ since $\delta_\times = 3$.
				Note that `digits' $z_j$ for $j < 0$ are ignored.
				$P$-digit online addition lies at the heart of the algorithm; due to its fixed width, hardware that implements Algorithm~\ref{alg:classic_om} can multiply to a precision of at most $P$ digits, which must be fixed in advance.
				
				\renewcommand{\algorithmicrequire}{\textbf{Inputs:}}
				\renewcommand{\algorithmicensure}{\textbf{Output:}}
				
				\begin{algorithm}[t]
					\begin{algorithmic}[1]
						\REQUIRE serially presented multiplicand $x$, multiplier $y$
						\STATE $\boldsymbol{x}, \boldsymbol{y}, \boldsymbol{w} \gets \boldsymbol{0}$
						\FOR{$j = 0$ \TO $P + 2$}
							\STATE $\boldsymbol{y} \gets \boldsymbol{y} \concat y_{j}$
							\STATE $\boldsymbol{v} \gets 2\boldsymbol{w} + 2^{-3}{\left(\boldsymbol{x}y_{j} + \boldsymbol{y}x_{j}\right)}$
							\STATE $z_{j-3} \gets \textnormal{sel}_\times{\left(\boldsymbol{v}\right)}$
							\STATE $\boldsymbol{w} \gets \boldsymbol{v} - z_{j-3}$
							\STATE $\boldsymbol{x} \gets \boldsymbol{x} \concat x_{j}$
						\ENDFOR
						\ENSURE serially generated product $z$
					\end{algorithmic}
					\caption{Radix-2 online multiplication~\cite{OAbook}.}
					\label{alg:classic_om}
				\end{algorithm}
			
			\subsubsection{Division}
			\label{subsubsec:od}
			
				The process of classical radix-2 online division is shown in Algorithm~\ref{alg:classic_od}, in which dividend $x$ and divisor $y$ are used to produce quotient $z$.
				In contrast to Algorithm~\ref{alg:classic_om}, division requires the formation of digit vector $\boldsymbol{z}$ since all prior output digits are needed for the calculation of $\boldsymbol{v}$, while updates to $\boldsymbol{w}$ require the full history of $y$.
				Online division therefore has more complex computation dependencies than multiplication.
				Its digit-selection function, $\textnormal{sel}_\div$, is
				\begin{equation*}
					\textnormal{sel}_\div{\left(\boldsymbol{v}\right)} =
					\begin{cases}
						1	&	\textnormal{if}~\boldsymbol{v} \geq \nicefrac{1}{4}						\\
						0	&	\textnormal{if}~-\nicefrac{1}{4} \leq \boldsymbol{v} < \nicefrac{1}{4}	\\
						-1	&	\textnormal{otherwise}.
					\end{cases}
				\end{equation*}
				$z_j$ is produced at cycle $j + 4$ since $\delta_\div$ is 4.
				
				\begin{algorithm}[t]
					\begin{algorithmic}[1]
						\REQUIRE serially presented dividend $x$, divisor $y$
						\STATE $\boldsymbol{y}, \boldsymbol{w}, \boldsymbol{z} \gets \boldsymbol{0}$
						\FOR{$j = 0$ \TO $P + 3$}
							\STATE $\boldsymbol{y} \gets \boldsymbol{y} \concat y_j$
							\STATE $\boldsymbol{v} \gets 2\boldsymbol{w} + 2^{-4}{\left(x_j - \boldsymbol{z}y_j\right)}$
							\STATE $z_{j-4} \gets \textnormal{sel}_\div{\left(\boldsymbol{v}\right)}$
							\STATE $\boldsymbol{w} \gets \boldsymbol{v} - z_{j-4}\boldsymbol{y}$
							\STATE $\boldsymbol{z} \gets \boldsymbol{z} \concat z_{j-4}$
						\ENDFOR
						\ENSURE serially generated quotient $z$
					\end{algorithmic}
					\caption{Radix-2 online division~\cite{OAbook}.}
					\label{alg:classic_od}
				\end{algorithm}
	
	\section{Proposed ARCHITECTure}
	\label{sec:architect}
		
		Using classical online operators as a starting point, we now describe the construction of constant-compute-resource hardware capable of performing iterative computation to increasing precision over time.
		We call this concept \architect{}.
		
		\subsection{Digit-vector Storage}
		\label{subsec:storage}
			
			Classical online operators make use of registers to store digit vectors.
			When implementing Algorithm~\ref{alg:classic_om} in hardware, for example, $P$-digit registers are needed for $\boldsymbol{x}$ and $\boldsymbol{y}$.
			To compute to an arbitrary precision $p$ instead, this is unsuitable; we must use random-access memory (RAM) for digit-vector storage to avoid both under- and over-budgeting register resources.
			We break $p$ into two dimensions: one fixed, $U$, that determines the RAM width, and a second variable, $n = \ceil*{\nicefrac{p}{U}}$, representing the number of these `chunks' that constitute each $p$-digit number.
			For digit index $i$, where $0 \leq i < p$, we define chunk index $c = \floor*{\nicefrac{i}{U}}$ and chunk digit index $u = i \bmod U$ such that $i = Uc + u$.
			When performing iterative calculations, independent digit vectors exist for each step, thus their indexing requires three variables: $c \in \left[0, n\right)$, $u \in \left[0, U\right)$ and approximant index $k$.
			
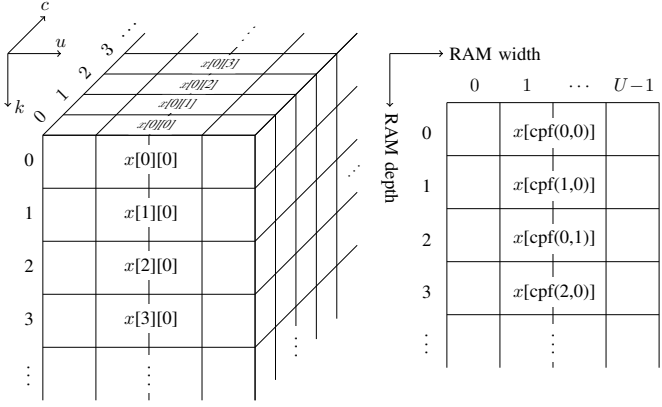
\begin{figure}
				\centering
				\resizebox{\columnwidth}{!}{
	\begin{tikzpicture}[yscale=1, xscale=1]
	\begin{scope}[yscale=1]
	\draw[->] (-2.2,4) -- (-1.2,4) node[above] {$u$};
	\draw[->] (-2.2,4) -- (-2.2,3) node[right] {$k$};
	\draw[->] (-2.2,4) -- (-1.5, 4.7) node[above] {$c$};
	
	\foreach \x in{0,...,4}
	{   \draw (0,\x ,4) -- (4,\x ,4); 
		\draw (\x ,-1,4) -- (\x ,4,4);
		\draw (4,\x ,4) -- (4,\x ,-1); 
		\draw (\x ,4,4) -- (\x ,4,-1);
		\draw (4,-1,\x ) -- (4,4,\x ); 
		\draw (0,4,\x ) -- (4,4,\x ); 
	}
 	\node[draw=none,fill=white] at (0.5,2) {$x$[0][0]}; 
 	\node[draw=none,fill=white] at (0.5,1) {$x$[1][0]}; 
 	\node[draw=none,fill=white] at (0.5,0) {$x$[2][0]}; 
 	\node[draw=none,fill=white] at (0.5,-1) {$x$[3][0]}; 
 	\node[draw=none,fill=white,rotate = 90] at (0.475,-2.05) {$\cdots$}; 
 	
 	\node[draw=none,fill=white,xslant=0.7,scale=0.65] at (0.6,2.65) { $x$[0][0]}; 
 	\node[draw=none,fill=white,xslant=0.7,scale=0.65] at (1.0,3.05) { $x$[0][1]};
 	\node[draw=none,fill=white,xslant=0.7,scale=0.65] at (1.4,3.42) { $x$[0][2]};
 	\node[draw=none,fill=white,xslant=0.7,scale=0.65] at (1.8,3.80) { $x$[0][3]};
 	\node[draw=none,fill=white,xslant=0.7, scale=0.8, rotate=80] at (2.37,4.4) { $\cdots$};
 	
	\node[draw=none,fill=white, rotate = 45] at (-0.7,3.7) {$ 0 \hspace{12pt} 1 \hspace{12pt} 2 \hspace{12pt}  3 \hspace{6pt}   \cdots$};  
 	\node[draw=none,fill=white] at (-1.8,2) {0}; 
	\node[draw=none,fill=white] at (-1.8,1) {1}; 
	\node[draw=none,fill=white] at (-1.8,0) {2}; 
	\node[draw=none,fill=white] at (-1.8,-1) {3}; 
	\node[draw=none,fill=white,rotate = 90] at (-1.8,-2.05) {$\cdots$}; 
 	
 	\node[draw=none,fill=white,rotate=90] at (3.245,-1.5) { $\cdots$};
 	\node[draw=none,fill=white,scale=0.8,rotate=45] at (4.35,1.8) { $\cdots$};
 	\end{scope}
 	
 	\begin{scope}[xshift=7cm, yscale=1]
 		
 		\draw[->] (-2,4) -- (-1,4) node[right] {RAM width};
 		\draw[->] (-2,4) -- (-2,3) node[right, rotate =-90] {RAM depth};
 		
 		\foreach \x in{1,...,5}
 		{   \draw (1,\x ,5) -- (5,\x ,5); 
 			\draw (\x ,0,5) -- (\x ,5,5);
 		}
 	
 		\node[draw=none,fill=white] at (1.25,3.4) {$ 0 \hspace{23pt} 1 \hspace{18pt} \cdots \hspace{11pt}  U\!-\!1$}; 
 		\node[draw=none,fill=white] at (1.1,2.5) {$x$[cpf(0,0)]}; 
 		\node[draw=none,fill=white] at (1.1,1.5) {$x$[cpf(1,0)]}; 
 		\node[draw=none,fill=white] at (1.1,0.5) {$x$[cpf(0,1)]}; 
 		\node[draw=none,fill=white] at (1.1,-0.5) {$x$[cpf(2,0)]};   
 		\node[draw=none,fill=white,rotate=90] at (1.09,-1.45) {$\cdots$}; 
	 	\node[draw=none,fill=white] at (-1.3,2.5) {0}; 
 		\node[draw=none,fill=white] at (-1.3,1.5) {1}; 
 		\node[draw=none,fill=white] at (-1.3,0.5) {2}; 
 		\node[draw=none,fill=white] at (-1.3,-0.5) {3}; 
 		\node[draw=none,fill=white,rotate = 90] at (-1.3,-1.45) {$\cdots$}; 
 		
 		\end{scope}
	\end{tikzpicture}
}	
				\caption{
					Operation of our Cantor pairing function, showing the transformation of a three-dimensional array growing with both approximant and chunk indices $k$ and $c$ to a structure growing only in a single dimension.
				}
				\label{fig:storage}
			\end{figure}

			Since \architect{} requires $k$ and $i$ to both vary non-monotonically as time progresses, it is necessary to uniquely encode a one-to-one mapping from two-dimensional approximant and chunk index pair $\left(k, c\right)$ into one-dimensional time.	
			We use a Cantor pairing function (CPF)~\cite{cegielski2001decidability}, a bijection from $\mathbb{N}^2$ onto $\mathbb{N}$, for this purpose, defined to be
			\begin{equation}
				\textnormal{cpf}{\left(k, c\right)} = \frac{\left(k + c\right)\left(k + c + 1\right)}{2} + c.
				\label{eq:cpf}
			\end{equation}
			The function's bijectivity is crucial for \architect{}.
			Unlike classical row- or column-major indexing, the injectivity of the CPF allows both dimensions to grow without bound while providing a unique result for every $\left(k, c\right)$.
			Its operation is demonstrated visually in Fig.~\ref{fig:storage}; what is conceptually a three-dimensional array indexed as $\left(k, c, u\right)$ becomes a two-dimensional array indexed by $\left(\textnormal{cpf}{\left(k, c\right)}, u\right)$ instead, thereby suiting the `flat' nature of RAM.
			The function's surjectivity ensures that every $\textnormal{cpf}{\left(k, c\right)}$ is produced by some $\left(k, c\right)$, thus enabling the most efficient use of the available memory.
		
		\vspace{-2mm}
		\subsection{Arbitrary-precision Operators}
		\label{sec:architect_op}
		
			\subsubsection{Multiplication}
			\label{subsec:architect_mul}
			
				We are now in a position to rewrite Algorithm~\ref{alg:classic_om} such that it can compute results to arbitrary precision.
				These transformed steps are shown in Algorithm~\ref{alg:ap_om}.
				Most importantly, a new loop has been introduced; this iterates over the $n$ pairs of $p$-digit numbers' chunks, most significant first, to facilitate arbitrary-precision multiplication with a $U$-digit online adder.   
				Digit vectors $\boldsymbol{x}$, $\boldsymbol{y}$, $\boldsymbol{v}$ and $\boldsymbol{w}$ are now indexed in two dimensions, corresponding to standard RAM addressing denoted as $\left[\textnormal{word}\right]\!\left[\textnormal{digit}\right]$.
				Where a digit index is not given, all $U$ digits of that word are accessed simultaneously.
				
				\begin{algorithm}[t]
					\begin{algorithmic}[1]
						\REQUIRE serially presented multiplicand $x$, multiplier $y$; approximant index $k$, precision $p$
						\STATE $\boldsymbol{x}, \boldsymbol{y}, \boldsymbol{w} \gets \boldsymbol{0}$
						\FOR{$j = 0$ \TO $p + 2$}
							\STATE $\boldsymbol{y}{\left[\textnormal{cpf}{\left(k, \floor*{\nicefrac{j}{U}}\right)}\right]\!\left[j \bmod U\right]} \gets y_{j}$
							\FOR{$c = \floor*{\nicefrac{j}{U}}$ \TO $0$}
								\STATE $\boldsymbol{v}{\left[\textnormal{cpf}{\left(k, c\right)}\right]} \gets 2\boldsymbol{w}{\left[\textnormal{cpf}{\left(k, c\right)}\right]} +$\\
								\quad $2^{-3}{\left(\boldsymbol{x}{\left[\textnormal{cpf}{\left(k, c\right)}\right]}y_{j} + \boldsymbol{y}{\left[\textnormal{cpf}{\left(k, c\right)}\right]}x_{j}\right)}$
								\IF{$c > 0$}
									\STATE $\boldsymbol{w}{\left[\textnormal{cpf}{\left(k, c\right)}\right]} \gets \boldsymbol{v}{\left[\textnormal{cpf}{\left(k, c\right)}\right]}$
								\ENDIF
							\ENDFOR
							\STATE $z_{j-3} \gets \textnormal{sel}_\times{\left(\boldsymbol{v}{\left[\textnormal{cpf}{\left(k, 0\right)}\right]}\right)}$
							\STATE $\boldsymbol{w}{\left[\textnormal{cpf}{\left(k, 0\right)}\right]} \gets \boldsymbol{v}{\left[\textnormal{cpf}{\left(k, 0\right)}\right]} - z_{j-3}$
							\STATE $\boldsymbol{x}{\left[\textnormal{cpf}{\left(k, \floor*{\nicefrac{j}{U}}\right)}\right]\!\left[j \bmod U\right]} \gets x_{j}$
						\ENDFOR
						\ENSURE serially generated product $z$
					\end{algorithmic}
					\caption{Radix-2 \architect{} multiplication.}
					\label{alg:ap_om}
				\end{algorithm}
			
			\subsubsection{Division}
			\label{subsec:architect_div}
			
				The equivalently transformed version of Algorithm~\ref{alg:classic_od} is shown in Algorithm~\ref{alg:ap_od}.
				Mirroring the increased complexity of classical online division over multiplication, here, two accumulation loops are needed: one for the calculation of $\boldsymbol{v}$, as for multiplication, and a second for $\boldsymbol{w}$.
				Consequently, $n - 1$ more cycles are required for the computation of an output digit in \architect{} division than multiplication.
				
				\begin{algorithm}[t]
					\begin{algorithmic}[1]
						\REQUIRE serially presented dividend $x$, divisor $y$; approximant index $k$, precision $p$
						\STATE $\boldsymbol{y}, \boldsymbol{w}, \boldsymbol{z} \gets 0$
						\FOR {$j = 0$ \TO $p + 3$}
							\STATE $\boldsymbol{y}{\left[\textnormal{cpf}{\left(k, \floor*{\nicefrac{j}{U}}\right)}\right]\!\left[j \bmod U\right]} \gets y_{j}$
							\FOR{$c = \floor*{\nicefrac{j}{U}}$ \TO $0$}
								\STATE $\boldsymbol{v}{\left[\textnormal{cpf}{\left(k, c\right)}\right]}\!\gets\!2\boldsymbol{w}{\left[\textnormal{cpf}{\left(k, c\right)}\right]} +$\\
								\quad $2^{-4}(x_{j} - \boldsymbol{z}{\left[\textnormal{cpf}{\left(k, c\right)}\right]}y_{j})$
							\ENDFOR
							\STATE $z_{j-4} \gets \textnormal{sel}_\div{\left(\boldsymbol{v}{\left[\textnormal{cpf}{\left(k, 0\right)}\right]}\right)}$
							\FOR {$c = \floor*{\nicefrac{j}{U}}$ \TO $0$}
								\STATE $\boldsymbol{w}{\left[\textnormal{cpf}{\left(k, c\right)}\right]} \gets \boldsymbol{v}{\left[\textnormal{cpf}{\left(k, c\right)}\right]} - z_{j-4}\boldsymbol{y}{\left[\textnormal{cpf}{\left(k, c\right)}\right]}$
							\ENDFOR
							\STATE $\boldsymbol{z}{\left[\textnormal{cpf}{\left(k, \floor*{\nicefrac{j}{U}}\right)}\right]\!\left[j \bmod U\right]} \gets z_{j-4}$
						\ENDFOR
						\ENSURE serially generated quotient $z$	
					\end{algorithmic}
					\caption{Radix-2 \architect{} division.}
					\label{alg:ap_od}
				\end{algorithm}
				
				Particular care is required for digit alignment in online division since input operands need to be bounded such that the output range is $\left(-1, 1\right)$~\cite{OAbook}.
				The normalisation of quotients following online division ordinarily necessitates variable $\delta_\div$~\cite{tu1989design}.
				To avoid this, we can maintain a fixed online delay by bounding divisor magnitude within $\left[\nicefrac{1}{r}, 1\right)$~\cite{obermann1997division}.
				For experimentation, we can guarantee alignment across iterations through the appropriate selection of initial inputs.
			
		\subsection{Digit Computation Scheduling}
		\label{subsec:alg_spec_patt}
			
			Given a generic online delay $\delta$ made up of latencies from a pipeline (or replicated pipelines operating in parallel) of one or more operators implementing the body of an iterative algorithm, restrictions are imposed on the order in which digits can be calculated.
			$\delta$ impacts us in two ways:
			\begin{itemize}
				\item Calculation of the first output digit requires the prior input of the first $\delta + 1$ input digits. Thereafter, each subsequent output digit requires one additional input digit in order to be computed.
				\item The $i^\textnormal{th}$ output digit is generated $\delta$ cycles after the $i^\textnormal{th}$ input digit is presented.
			\end{itemize}
			In general, digits of the same approximant can be calculated indefinitely, while those across iterations must be sequenced such that they obey these $\delta$-imposed limitations.
			When scheduling digit $z_i\iter{k}$'s generation, we must ensure that
			\begin{equation*}
				t{\left(z_{i + 1}\iter{k}\right)} > t{\left(z_i\iter{k}\right)}, \quad
				t{\left(z_{i}\iter{k + 1}\right)} > t{\left(z_{i + \delta}\iter{k}\right)}
			\end{equation*}
			for all approximant indices $k \ge 1$ and digit indices $i \ge 0$, where $t$ is the time at which a generation event occurs.
			
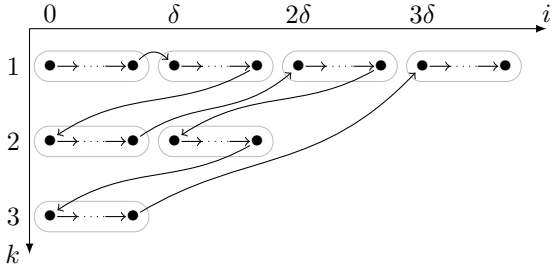
\begin{figure}
				\centering
				\begin{tikzpicture}[yscale=-1, xscale=0.55]
	
	\def\xz{-0.3}
	\def\xa{1.0}
	\def\xn{3.3}
	\def\xb{4.3}
	\def\barh{-2.5 mm}
	\def\bari{15 mm}
	\def\barb{19 mm}
	\def\barj{22 mm}
	\def\barc{25 mm}
	\tikzset{  
		brace/.style={decorate, decoration={brace,amplitude=1mm}},
	}
	
	\node[] at (0,-0.7) {$0$};
	\node[] at (3,-0.7) {$\delta$};
	\node[] at (6,-0.7) {$2\delta$};
	\node[] at (9,-0.7) {$3\delta$};
	\node[] at (12,-0.7) {$i$};
	\node[anchor=east] at (-0.5, 0) {$1$};
	\node[anchor=east] at (-0.5, 1) {$2$};
	\node[anchor=east] at (-0.5, 2) {$3$};
	\foreach \i/\j in {
		0/0, 2/0, 3/0, 5/0, 6/0, 8/0, 9/0, 11/0,
		0/1, 2/1, 3/1, 5/1,
		0/2, 2/2, 
	} {
		\node[inner sep=0] (n\i\j) at (\i,\j) {$\bullet$};
		\node[white] (n11) at (1,1) {\large$\bullet$};
	}
	
	\draw[-latex] (-0.5,-0.5) to (12,-0.5);
	\draw[-latex] (-0.5,-0.5) to (-0.5,2.5);
	\node[anchor=east] at (-0.5,2.5) {$k$};
	
	\newcommand\braceheight{-0.6}
	
%
%
	
	\foreach \i/\j [
	evaluate=\i as \ia using \i+(2/3),
	evaluate=\i as \ib using \i+2-(2/3),
	evaluate=\i as \ic using int(\i+2)
	] in {0/0, 0/1, 0/2} {
		\draw[->] (n\i\j) -- (\ia,\j);
		\draw[dotted] (\ia,\j) -- (\ib,\j);
		\draw[->] (\ib,\j) -- (n\ic\j);
		\node[draw=black!25, rounded corners=2mm, fit=(n\i\j)(n\ic\j)] (blob\i\j) {};
	}

	\foreach \i/\j [
	evaluate=\i as \ia using \i+(2/3),
	evaluate=\i as \ib using \i+2-(2/3),
	evaluate=\i as \ic using int(\i+2)
	] in {3/0, 3/1, 6/0, 9/0} {
		\draw[->] (n\i\j) -- (\ia,\j);
		\draw[dotted] (\ia,\j) -- (\ib,\j);
		\draw[->] (\ib,\j) -- (n\ic\j);
		\node[draw=black!25, rounded corners=2mm, fit=(n\i\j)(n\ic\j)] (blob\i\j) {};
	}
	
	\draw[->] (n20) to[bend right] (n30);
	\draw[->] (n50) to[out=180-20, in=360-20] (n01.north east);
	\draw[->] (n21) to[out=360-25, in=180-25] (n60.south west);
	\draw[->] (n80) to[out=180-20, in=360-20] (n31.north east);
	\draw[->] (n51) to[out=180-20, in=360-20] (n02.north east);
	\draw[->] (n22) to[out=-20, in=150] (n90.south west);
	
\end{tikzpicture}
				\caption{Proposed digit generation pattern without don't-change digit elision for generic iterative computation using online operators.}
				\label{fig:digit_pattern_TC}
			\end{figure}

			While we have the freedom to trade off between iteration count and precision within the bounds of these dependencies, we always assume a mapping from current to next digit of the form depicted in Fig.~\ref{fig:digit_pattern_TC}.
			The groups of digits shown, each $\delta$ in size, are processed `downwards' and `leftwards,' with slope dependent on $\delta$ and control snapping back to the first approximant once digit position $i = 0$ has been reached.
			Fixing the granularity of digit generation to $\delta$ allows for control path simplification---as will be elaborated upon in Section~\ref{subsec:control}---and limits transitions between approximants.
			The latter is beneficial since, as will be explained in Section~\ref{subsubsec:comp_time_in_clk}, switching between approximants leads to the incursion of performance penalties under some circumstances.
			
		\subsection{Don't-change Digit Elision}
		\label{subsec:dontchange}
			
			Thanks to the use of online arithmetic, when advancing downwards in our iteration-precision space, we can avoid the recalculation of don't-change digits, \textit{i.e.} those of later approximants that have stabilised.
			This is generally not possible in LSD-first architectures, in which carries can propagate from LSD to MSD.
			Don't-change digit elision is guaranteed to be an error-free transformation: it induces no approximation.
			
			The concept behind this optimisation is straightforward.
			Before beginning to calculate the digits of approximant $k$, we examine the digits of the previous two approximants.
			If these approximants are equal in their most-significant $q + \delta$ digits, it is guaranteed that approximant $k$ will be equal to its two predecessors in its first $q$ digits.
			Hence, we do not need to calculate them; we can skip directly to digit $q$'s generation.
			
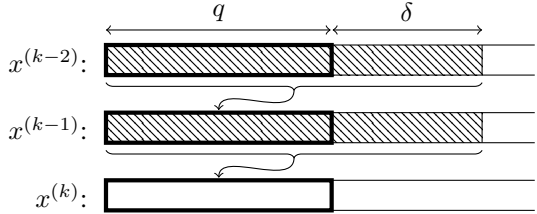
\begin{figure}
				\centering
				\begin{tikzpicture}[yscale=-1]

	\def\xz{0}
	\def\xn{3}
	\def\xd{5}
	\def\xm{5.7}
	\def\barh{4mm}
	\def\bari{9mm}

	\tikzset{  
	  brace/.style={decorate, decoration={brace,amplitude=1mm}},
	}

	\draw[<->, shorten >=0.4pt] (\xz, -2mm) to [auto] node {$q$} (\xn, -2mm); 
	\draw[<->, shorten <=0.4pt] (\xn, -2mm) to [auto] node {$\delta$} (\xd, -2mm); 

	\draw (\xz, 0*\bari) -- (\xm, 0*\bari);
	\draw (\xz, 0*\bari+\barh) -- (\xm, 0*\bari+\barh);
	\draw (\xz, 0*\bari) to [auto,swap] node {$x\iter{k-2}$:~~} (\xz, 0*\bari+\barh);
	\draw (\xn, 0*\bari) -- (\xn, 0*\bari+\barh);
	\draw (\xd, 0*\bari) -- (\xd, 0*\bari+\barh);

	\draw[brace] (\xd, 0*\bari+\barh+1mm) to coordinate (p0) (\xz, 0*\bari+\barh+1mm);
	\draw[->, shorten >=1pt]([yshift=1mm]p0) to [out=90, in=270] (0.5*\xz+0.5*\xn, 1*\bari);

	\draw (\xz, 1*\bari) -- (\xm, 1*\bari);
	\draw (\xz, 1*\bari+\barh) -- (\xm, 1*\bari+\barh);
	\draw (\xz, 1*\bari) to [auto,swap] node {$x\iter{k-1}$:~~} (\xz, 1*\bari+\barh);
	\draw (\xn, 1*\bari) -- (\xn, 1*\bari+\barh);
	\draw (\xd, 1*\bari) -- (\xd, 1*\bari+\barh);

	\draw[brace] (\xd, 1*\bari+\barh+1mm) to coordinate (p0) (\xz, 1*\bari+\barh+1mm);
	\draw[->, shorten >=1pt]([yshift=1mm]p0) to [out=90, in=270] (0.5*\xz+0.5*\xn, 2*\bari);

	\draw (\xz, 2*\bari) -- (\xm, 2*\bari);
	\draw (\xz, 2*\bari+\barh) -- (\xm, 2*\bari+\barh);
	\draw (\xz, 2*\bari) to [auto,swap] node {$x\iter{k}$:~~} (\xz, 2*\bari+\barh);
	\draw (\xn, 2*\bari) -- (\xn, 2*\bari+\barh);

	\draw[draw=none, pattern=north west lines] (\xz, 0*\bari) rectangle (\xd, 0*\bari+\barh);
	\draw[draw=none, pattern=north west lines] (\xz, 1*\bari) rectangle (\xd, 1*\bari+\barh);

	\draw[ultra thick] (\xz, 0*\bari) rectangle (\xn, 0*\bari+\barh);
	\draw[ultra thick] (\xz, 1*\bari) rectangle (\xn, 1*\bari+\barh);
	\draw[ultra thick] (\xz, 2*\bari) rectangle (\xn, 2*\bari+\barh);

\end{tikzpicture}
				\caption{
					A proof sketch showing why it is sound to omit don't-change digits.
					If the two hatched regions contain the same $q + \delta$ digits, the three thick boxes are guaranteed to contain the same $q$ digits, hence $x\iter{k}$'s calculation can begin from digit index $q$.
				}
				\label{fig:dont_change}
			\end{figure}

			The soundness of this optimisation can be justified by appealing to the digit dependencies of online arithmetic.	
			Fig.~\ref{fig:dont_change} provides some graphical intuition.
			Given that each approximant depends only on the value of its immediate predecessor, and recalling the definition of online delay from Section~\ref{subsec:online}, we emphasise that the first $q$ digits of one approximant depend only upon the first $q + \delta$ digits of the previous approximant~\cite{OAbook}.
			Hence, if approximants $k - 2$ and $k - 1$ are equal in their first $q + \delta$ digits, approximant $k$ is guaranteed to be equal to them in its first $q$ digits.
			
			During the generation of approximant $k$, we compare digits on the fly with those generated for approximant $k - 1$, previously stored in RAM.
			Based on the number of digits found to be equal, we store a pointer indicating whence approximant $k + 1$'s, \emph{i.e.} the \emph{next} approximant's, generation should begin.
			Pointer storage requires a small amount of extra memory but, as will be elaborated upon in Section~\ref{subsec:performance_composite}, this overhead is small and amortised out the more RAM is instantiated for storing digit vectors.
			Since we have elected to process digits in groups of $\delta$, it makes sense to also limit our don't-change digit elision to this granularity.
			We thus avoid the processing of entire groups of digits, where possible.
			
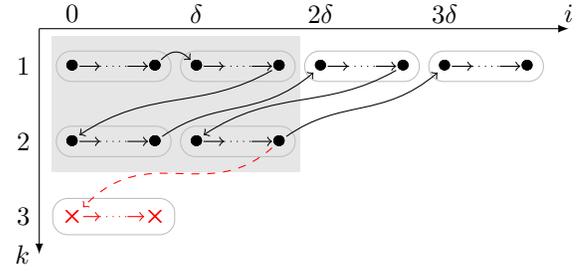
\begin{figure}
				\centering
				\begin{tikzpicture}[yscale=-1, xscale=0.55]
	
	\def\xz{-0.3}
	\def\xa{1.0}
	\def\xn{3.3}
	\def\xb{5.3}
	\def\barh{-2.5 mm}
	\def\bari{15 mm}
	\def\barb{19 mm}
	\def\barj{22 mm}
	\def\barc{25 mm}
	\tikzset{  
		brace/.style={decorate, decoration={brace,amplitude=1mm}},
	}
	
	\foreach \i/\j in {
		6/0, 8/0, 9/0, 11/0,		
	} {
		\node[inner sep=0] (n60) at (6,0) {$\bullet$};
		\node[inner sep=0] (n80) at (8,0) {$\bullet$};
		\node[inner sep=0] (n90) at (9,0) {$\bullet$};
		\node[inner sep=0] (n110) at (11,0) {$\bullet$};
	}

	\foreach \i/\j in {
		0/0, 2/0, 3/0, 5/0, 
		0/1, 2/1, 3/1, 5/1,	
	} {
		\node[inner sep=0] (n00) at (0,0) {$\bullet$};
		\node[inner sep=0] (n20) at (2,0) {$\bullet$};
		\node[inner sep=0] (n30) at (3,0) {$\bullet$};
		\node[inner sep=0] (n50) at (5,0) {$\bullet$};
		\node[inner sep=0] (n01) at (0,1) {$\bullet$};
		\node[inner sep=0] (n21) at (2,1) {$\bullet$};
		\node[inner sep=0] (n31) at (3,1) {$\bullet$};
		\node[inner sep=0] (n51) at (5,1) {$\bullet$};
	}

	\foreach \i/\j in {
	0/2, 2/2,
	} {
		\node[red, inner sep=0] (n02) at (0,2) {$\times$};
		\node[red, inner sep=0] (n22) at (2,2) {$\times$};
	}	

	\draw[-latex] (-0.8,-0.5) to (12,-0.5);
	\draw[-latex] (-0.8,-0.5) to (-0.8,2.5);
	\node[] at (12,-0.7) {$i$};
	\node[anchor=east] at (-0.8,2.5) {$k$};
	\node[] at (0,-0.7) {$0$};
	\node[] at (3,-0.7) {$\delta$};
	\node[] at (6,-0.7) {$2\delta$};
	\node[] at (9,-0.7) {$3\delta$};
	\node[anchor=east] at (-0.8, 0) {$1$};
	\node[anchor=east] at (-0.8, 1) {$2$};
	\node[anchor=east] at (-0.8, 2) {$3$};
	\newcommand\braceheight{-0.6}
	
%
%
	
	\foreach \i/\j [
	evaluate=\i as \ia using \i+(2/3),
	evaluate=\i as \ib using \i+2-(2/3),
	evaluate=\i as \ic using int(\i+2)
	] in {0/0, 0/1} {
		\draw[->] (n\i\j) -- (\ia,\j);
		\draw[dotted] (\ia,\j) -- (\ib,\j);
		\draw[->] (\ib,\j) -- (n\ic\j);
		\node[draw=black!25, rounded corners=2mm, fit=(n\i\j)(n\ic\j)] (blob\i\j) {};
	}

	\foreach \i/\j [
	evaluate=\i as \ia using \i+(2/3),
	evaluate=\i as \ib using \i+2-(2/3),
	evaluate=\i as \ic using int(\i+2)
	] in {0/2} {
		\draw[red, ->] (n\i\j) -- (\ia,\j);
		\draw[red, dotted] (\ia,\j) -- (\ib,\j);
		\draw[red, ->] (\ib,\j) -- (n\ic\j);
		\node[draw=black!25, rounded corners=2mm, fit=(n\i\j)(n\ic\j)] (blob\i\j) {};
	}
	
	\foreach \i/\j [
	evaluate=\i as \ia using \i+(2/3),
	evaluate=\i as \ib using \i+2-(2/3),
	evaluate=\i as \ic using int(\i+2)
	] in {3/0, 3/1, 6/0,9/0} {
		\draw[->] (n\i\j) -- (\ia,\j);
		\draw[dotted] (\ia,\j) -- (\ib,\j);
		\draw[->] (\ib,\j) -- (n\ic\j);
		\node[draw=black!25, rounded corners=2mm, fit=(n\i\j)(n\ic\j)] (blob\i\j) {};
	}
	
	\draw[->] (n20) to[bend right] (n30);
	\draw[->] (n50) to[out=180-20, in=360-20] (n01.north east);
	\draw[->] (n21) to[out=360-25, in=180-25] (n60.south west);
	\draw[->] (n80) to[out=180-20, in=360-20] (n31.north east);
	\draw[red, dashed,->] (n51) to[out=180-30, in=360-30] (n02.north east);
	\draw[->] (n51) to[out=360-25, in=180-25] (n90.south west);
	

	
	
\begin{scope}[on background layer]

\draw [draw=none, fill=black!10] ($(-0.5, -0.4)$) -- ($(5.5, -0.4)$) -- ($(5.5, 1.4)$) -- ($(-0.5, 1.4)$) -- ($(-0.5, -0.4)$);

\end{scope}
	
	
\end{tikzpicture}
				\caption{
					Digit generation pattern with don't-change digit elision.
					Groups of digits in the shaded region were found to be identical at runtime, allowing computation of the first group to be skipped in the subsequent iteration.
					Dashed lines are scheduled paths not taken and $\times$s are digits therefore elided.
				}
				\label{fig:digit_pattern_msd}
			\end{figure}

			As a result of the introduction of don't-change digit elision, our scheduling pattern becomes dynamic.
			Fig.~\ref{fig:digit_pattern_msd} shows an example.
			This is similar to Fig.~\ref{fig:digit_pattern_TC}, but, due to the identification of the third approximant's first group of MSDs as stable, we can advance into the iteration-precision space more quickly than had we not elided them, increasing compute efficiency.

			Along with increased performance, the elision of don't-change digits also enables us to increase memory efficiency.
			Defining $\psi$ as the number of digits guaranteed not to have changed within the current approximant, as determined through the runtime comparison of MSDs within the preceding two approximants, we can substitute
			\begin{equation*}
				\textnormal{cpf}{\left(k, \hat{c}\right)} = \frac{\left(k + \hat{c}\right)\left(k + \hat{c} + 1\right)}{2} + \hat{c},
			\end{equation*}
			for \eqref{eq:cpf}, where $\hat{c} = \floor{\nicefrac{\left(i - \psi + 1\right)}{U}}$.
			By doing so, stable digits no longer need to be recomputed or stored.
			In common with its predecessor, this optimised storage strategy guarantees no memory wastage through the surjectivity of its mapping from approximant and chunk indices to memory addresses.

		\subsection{Control Logic}
		\label{subsec:control}	
			
\begin{figure*}
				\centering
				\begin{tikzpicture}[line width=1pt, inner sep=1pt]

\node(ST) at (-3.5,-0.6) {Start};
\node[state, draw, line width=1pt, shape=circle, minimum width=2cm, align=center, inner sep=0.1cm] (q1) at (0,1) {Digit\\generation};
\node[state, right of=q1, draw, line width=1pt, shape=circle, minimum width=2cm, align=center, inner sep=0.1cm] (q2) at (5,1) {Accumulation};

\draw
(ST) edge[->, bend right, above] node[auto, pos=0.7]{$/~k \leftarrow 1,~i \leftarrow 0$} (q1)
(q1) edge[->, loop above] node[outer sep=5pt]{$i \bmod \delta=\delta - 1 \wedge i \geq \delta~$\fbox{$\wedge~i - \psi \neq \delta - 1$}$~/~i\leftarrow i - 2\delta + 1,~k\leftarrow k + 1$} (q1)
(q1) edge[->, loop left] node[outer sep=5pt]{$i \bmod \delta \neq \delta-1~/~i \leftarrow i + 1$} (q1)
(q1) edge[->, loop below] node[outer sep=5pt]{$i = \delta - 1~$\fbox{$\vee \left(i \bmod \delta = \delta - 1 \wedge i \geq \delta \wedge i - \psi = \delta - 1 \right)$}$~/~k \leftarrow 1,~i\leftarrow i + k\delta + 1$} (q1)
(q2) edge[->, loop right] node[outer sep=5pt]{$\gamma > 0~/~\gamma \leftarrow \gamma - 1$} (q2)
(q1) edge[->, bend left, above] node[outer sep=5pt]{$i \geq U~/~\gamma \leftarrow \alpha\floor{\nicefrac{i}{U}} - 1$} (q2)
(q2) edge[->, bend left, below] node[outer sep=5pt]{$\gamma = 0$} (q1);
\end{tikzpicture}
				\caption{
					FSM for digit computation scheduling.
					Transition edges are labelled with conditions and actions separated by slashes ($/$).
					If the datapath consists only of adders, the accumulation state is never entered.
					Otherwise, $\alpha = 2$ if the datapath contains one or more dividers, and is 1 in all other cases.
					Boxed conditions apply only when don't-change digit elision is active; they are otherwise ignored.
					Termination occurs either on demand or following memory exhaustion.
				}
				\label{fig:FSM_control}
			\end{figure*}
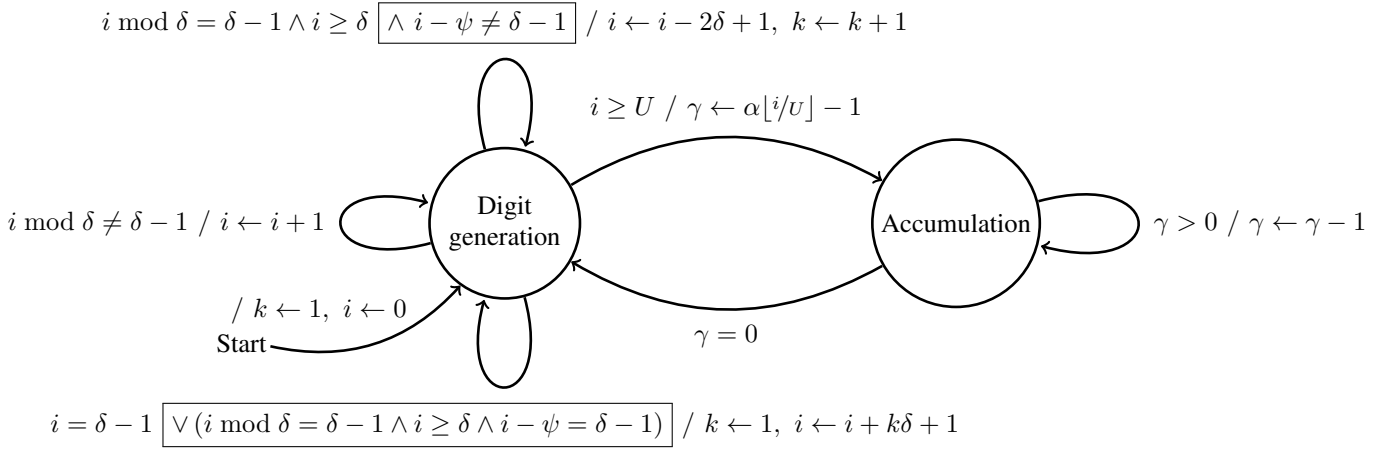

			Given a particular $\left(k, i\right)$, we can compute the subsequent $\left(k, i\right)$, $\left(k', i'\right)$, needed to realise scheduling patterns such as those shown in Figs~\ref{fig:digit_pattern_TC} and \ref{fig:digit_pattern_msd} with the finite-state machine (FSM) depicted in Fig.~\ref{fig:FSM_control}.
			Therein, we present state transition conditions both with and without don't-change digit elision functionality.
			When elision is enabled, the conditions shown in boxes are evaluated in addition to those outside.
			
			The states' functionality is as follows.
			\begin{itemize}
				\item \emph{Digit generation}:
					Manages the propagation and storage of $\delta$-digit groups across iterations.
					When remaining within this state, only digit index $i$ must be evaluated to determine changes needed to $k$ and $i$ without don't-change digit elision.
					When enabled, $\psi$ must also be considered.
				\item \emph{Accumulation}:
					Assuming that the constructed datapath contains at least one multiplier or divider, we must account for the variable latency of those operators.
					The throughput of the datapath as a whole is determined by the slowest operator.
					Since \architect{}'s multiplication and division operators have dissimilar accumulation functionality, as was explained in Section~\ref{sec:architect_op}, the number of clock cycles consumed by each is different.
					If the datapath contains at least one divider, advancement must be inhibited for $2\floor*{\nicefrac{i}{U}} - 1$ cycles per generated digit.
					If it does not, but does contain at least one multiplier, this factor is $\floor*{\nicefrac{i}{U}} - 1$ instead.
					Counter $\gamma$ sequences the return to the digit generation state.
					Since $i$ is variable, this loop cannot be unrolled.
					In the case that the datapath contains only adders, entry into this state never occurs.
			\end{itemize}
	
		\subsection{Accuracy Bounds}
		\label{subsec:kp_res}
			
			Let us assume the existence of a target result defined by its approximant index and precision $\left(K, P\right)$.
			To reach it, we are required to compute for at least $K$ iterations and to at least $P$-digit precision.
			We emphasise that \architect{} does not necessitate its users to specify $K$ or $P$ up-front, while other approaches require either one or both of these---usually $P$---to be determined before beginning to iterate.
			Since don't-change digits are identified at runtime, the analysis herein applies to \architect{} without digit elision.
			Enabling this optimisation will therefore increase the bounds that follow.
			
\begin{figure}
				\centering
				\begin{tikzpicture}[yscale=-1, xscale=0.8]

\def\xz{-0.3}
\def\xa{1.0}
\def\xn{3.3}
\def\xb{4.3}
\def\barh{-2.5 mm}
\def\bari{15 mm}
\def\barb{19 mm}
\def\barj{22 mm}
\def\barc{25 mm}
\tikzset{  
	brace/.style={decorate, decoration={brace,amplitude=1mm}},
}

\node[inner sep=0] (-0.5,3) at (-0.5,3) (K1) {$\bullet$};
\node[inner sep=0] (-0.5,2.5) at (-0.5,2.5) (K2) {$\bullet$};
\node[inner sep=0] (-0.5,1.15) at (-0.5,1.15) (K3) {$\bullet$};
\node[inner sep=0] (7,-0.5) at (7,-0.5) (P1) {$\bullet$};
\node[inner sep=0] (6,-0.5) at (6,-0.5) (P2) {$\bullet$};
\node[inner sep=0] (2.5,-0.5) at (2.5,-0.5) (P3) {$\bullet$};

\draw[-latex] (-0.5,-0.5) to (8,-0.5);
\node[anchor=south west] at (8,-0.5) {$i$};

\newcommand\braceheight{-0.6}

\node(Kr) at ($(P3.north)+(0, -0.2)$) {$P$};
\node(Pr) at ($(P2.north)+(0, -0.2)$) {$P_\textnormal{res}$};
\node(Pm) at ($(P1.north)+(0, -0.2)$) {$P_\textnormal{max}$};

\draw[->] (-0.5, -0.5) -- (-0.5, 1.15) node[left] {$K$} -- (-0.5, 2.5) node[left] {$K_\textnormal{res}$} -- (-0.5, 3) node[left] {$K_\textnormal{max}$} -- (-0.5, 3.5) node[left] {$k$};

\draw[-, dashed] (2.5, -0.5) to (2.5,1.15);
\draw[-, dashed] (-0.5, 1.15) to (2.5,1.15);
\node[blue] (2.5,1.13) at (2.5,1.13) (Target) {$\times$};
\node(Km) at ($(Target.west)+(-0.4, -0.3)$) {Target};

\draw[-] (7, -0.5) to (-0.5,3);
\draw[-,dashed] (6, -0.5) to (-0.5,2.5);
\end{tikzpicture}
				\caption{How the final precision and iteration count ($K_\textnormal{res}, P_\textnormal{res}$) are constrained by the desired result ($K, P$) and the available memory ($K_\textnormal{max}, P_\textnormal{max}$).}
				\label{fig:K_res}
			\end{figure}
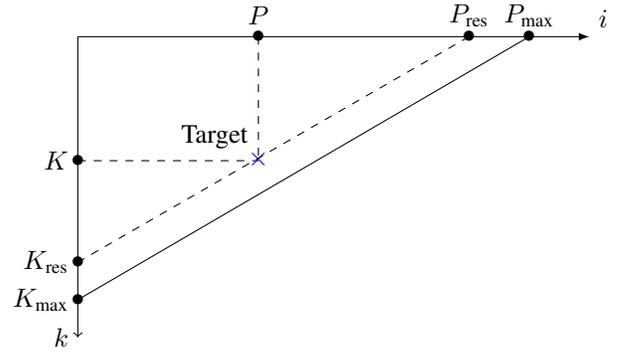

			As shown in Fig.~\ref{fig:K_res}, we define the number of iterations resulting from computation to target $\left(K, P\right)$ as $K_\textnormal{res}$ and the final precision of the first approximant---always the most precise---as $P_\textnormal{res}$.
			$K_\textnormal{res}$ is bounded to no more than $K_\textnormal{max}$, while $P_\textnormal{res}$ is similarly bounded by $P_\textnormal{max}$, both of which are determined by the size of the available memory.
			The latter therefore determines the maximum approximant index and precision---and consequently accuracy---that can be reached through the use of our approach.
			Thus, if higher accuracy is required, more memory must be instantiated.
			
			Upon termination, the precision of approximant $k$ will be
			\begin{equation*}
				p\iter{k} =
				\begin{cases}
					\delta{\left(\ceil*{\frac{P}{\delta}} + K - k\right)}	&	\textnormal{if}~k < K 	\\
					P														&	\textnormal{if}~k = K 	\\
					\delta{\left(K_\textnormal{res} - k \right)} 			&	\textnormal{otherwise},
				\end{cases}
			\end{equation*}
			where $K_\textnormal{res}$ can be geometrically deduced to be
			\begin{equation*}
				K_\textnormal{res} =
				\begin{cases}
					\ceil*{\frac{P}{\delta}} + K - 1	&	\textnormal{if}~P > \delta	\\
					K									&	\textnormal{otherwise}
				\end{cases}
			\end{equation*}
			and $P_\textnormal{res} = p\iter{1}$.
			
			For each arbitrary-precision digit vector to be stored, $K_\textnormal{max}$ and $P_\textnormal{max}$ are fixed by RAM depth $D$ (in $U$-digit words). 
			Analysis of our pairing function in \eqref{eq:cpf} allows us to derive
			\begin{align*}
				P_\textnormal{max}	&= U{\left(1 + \floor*{\nicefrac{3}{2}{\left(\sqrt{1 + \nicefrac{8}{9}D} - 1\right)}}\right)},	\\
				K_\textnormal{max}	&=
				\begin{cases}
					\frac{P_\textnormal{max}}{U} + 1	&	\textnormal{if}~D \geq {\left(\frac{P_\textnormal{max}}{U} + 1\right)}\frac{P_\textnormal{max}}{2U}  \\
					\frac{P_\textnormal{max}}{U}		&	\textnormal{otherwise}.
				\end{cases}
			\end{align*}
	
		\subsection{Compute Time}
		\label{subsubsec:comp_time_in_clk}
		
			Given a particular target $\left(K, P\right)$, and hence a certain $K_\textnormal{res}$ and $P_\textnormal{res}$, we can calculate the number of clock cycles required to compute the desired result.
			Let us first assume that don't-change digit elision is disabled.
			This total time $T$ can be broken down into the following three components such that $T = T_1 + T_2 + T_3$.
			\begin{itemize}
				\item \emph{Initial online delay}:
					We must wait $\delta$ clock cycles before each approximant's result begins to appear, thus the delay across all iterations is simply
					\begin{equation*}
						T_1 = \delta K_\textnormal{res}.
					\end{equation*}
				\item \emph{Digit generation}:
					Across all iterations performed, the total time for digit generation is either
					\begin{equation*}
						T_2 = \sum_{k = 0}^{K_\textnormal{res}-1}{p\iter{k}{\left(2n\iter{k} - 1\right)} - U{n\iter{k}}{\left(n\iter{k} - 1\right)}} - \delta,
					\end{equation*}
					if the datapath contains one or more dividers, or
					\begin{equation*}
						T_2 = \sum_{k = 0}^{K_\textnormal{res}-1}{n\iter{k}{\left(p\iter{k} - \frac{U{\left(n\iter{k} - 1\right)}}{2}\right)}} - \delta 
					\end{equation*}
					if it contains one or more multipliers.
					$n\iter{k} = \ceil*{\nicefrac{p\iter{k}}{U}}$ and represents the number of chunks within the given approximant upon termination of the algorithm.
					In the case that the datapath contains only one or more adders,
					\begin{equation*}
						T_2 = \sum_{k = 0}^{K_\textnormal{res}-1}{p\iter{k}} - \delta.
					\end{equation*}
					$p\iter{0}$ and $n\iter{0}$ are the numbers of digits and chunks, respectively, that must be read from the initial guess.
				\item \emph{Digit-serial addition}:
					Recall that a serial online adder has $\delta_+ = 2$. 
					When switching between iterations, adders, if present, require two cycles to recalculate the preceding approximant's residuals in order to produce a new digit~\cite{OAbook}.
					This ensures that the calculated digit aligns with its truncated digit vectors.
					For this,
					\begin{equation*}
						T_3 = \beta{\left(K_\textnormal{res}^2 - K_\textnormal{res} + 2K - 2\right)},
					\end{equation*}
					where $\beta$ is the number of serial adders present along the highest-online delay path within the circuit.
			\end{itemize}
			
			When enabled, don't-change digit elision generally allows computation time to be reduced below $T$.
			Since the amount of achievable reduction is input-dependent, however, it is not practicable to determine such reductions analytically.
			
		\subsection{Digit-parallel Addition Optimisation}
		
			It is possible to eliminate the final $T$ component in Section~\ref{subsubsec:comp_time_in_clk}, resulting in $T_3 = 0$, by using three-digit parallel online adders in place of serial ones.
			We store consecutive digit-vector words in alternating memory banks for speed.
			By ensuring that RAM width $U > 1$, \emph{i.e.} that each word contains at least two digits, we can always read the three contiguous digits required by these adders in a single cycle.
			No additional memory is needed for this optimisation.
	
	\section{Benchmarks}
	\label{sec:benchmark}
	
		In order to evaluate \architect{}, we implemented two widely used iterative algorithms---the Jacobi method (to solve systems of linear equations) and Newton's method (for the solution of nonlinear equations)---in hardware following the aforementioned principles.
		We chose Jacobi and Newton to exemplify a large class of iterative methods with linear and quadratic convergence properties, respectively.
		Except where otherwise stated, \architect{} implementations featured all of the previously described optimisations: don't-change digit elision, its related memory-addressing and digit-scheduling schemes and serial-to-parallel online adder substitutions.
	
		\subsection{Jacobi Method}
		\label{subsec:jacobi}		
				
			The Jacobi method seeks to solve the system of $N$ linear equations $\boldsymbol{A}\boldsymbol{x} = \boldsymbol{b}$.
			If $\boldsymbol{A}$ is decomposed into diagonal and remainder components such that $\boldsymbol{A} = \boldsymbol{D} + \boldsymbol{R}$, $\boldsymbol{x}$ can be computed through the repeated evaluation of
			\begin{equation*}
				\boldsymbol{x}\iter{k + 1} = \boldsymbol{D}^{-1}{\left(\boldsymbol{b} - \boldsymbol{R}\boldsymbol{x}\iter{k}\right)},
			\end{equation*}
			or, expressed in element-wise fashion,
			\begin{equation*}
				x_i\iter{k + 1} = \frac{1}{a_{ii}}{\left(b_i - \sum_{j \ne i \in \left[0, N\right)}{a_{ij}x_j\iter{k}}\right)}~\forall i \in \left[0, N\right),
			\end{equation*}
			where $k$ is the approximant index.
			Since $\boldsymbol{D}$'s only non-zero elements lie along its diagonal, $\boldsymbol{D}^{-1}$ is trivial to calculate.
			Note that $\boldsymbol{x}\iter{k + 1}$ relies only upon the previously computed value of $\boldsymbol{x}$; the calculation can therefore be parallelised by computing each $x_i\iter{k + 1}$ independently.
			A convergence criterion, $\norm{\boldsymbol{A}\boldsymbol{x}\iter{k} - \boldsymbol{b}} < \eta$, can be used in order to determine if the solution has been found to great enough accuracy.
			
			Such a system is guaranteed to be soluble when $\boldsymbol{A}$ is strictly diagonally dominant, \emph{i.e.} if the condition $\abs{a_{ii}} > \sum_{j \ne i}{\abs{a_{ij}}}$ holds for all $i$.
			Although strict diagonal dominance is not a necessity in every case, we assume this condition to always be satisfied for simplicity.
			
			A metric used to quantify the sensitivity of a particular linear system to error is the \emph{condition number} of $\boldsymbol{A}$~\cite{k_prec}, where
			\begin{equation*}
				\kappa{\left(\boldsymbol{A}\right)} = \left\|\boldsymbol{A}\right\|{\left\|\boldsymbol{A}^{-1}\right\|}.
			\end{equation*}
			Perturbations in $\boldsymbol{x}\iter{k}$, caused by rounding, lead to errors in $\boldsymbol{x}\iter{k + 1}$ whose magnitude is dependent, in part, on $\kappa{\left(\boldsymbol{A}\right)}$; a high condition number indicates that $\boldsymbol{A}$ is sensitive to error and therefore ill-conditioned~\cite{multiquadric}.		
			We can expect to need at least $\omega$ additional digits of precision in order to compute a system with $\kappa{\left(\boldsymbol{A}\right)} = 2^\omega$ than required if $\kappa{\left(\boldsymbol{A}\right)}$ were 1~\cite{numerical_book}.
			
\begin{figure}
				\centering
				\begin{subfigure}{\columnwidth}
					\centering
								\newcommand\busmark[3][west] {
				\draw[line width=1pt] (#2-0.1,#3-0.1) to (#2+0.1,#3+0.1);
				\node[anchor=#1] at (#2,#3) {\scriptsize 2};
			}
			\newcommand\busmarkp[3][west] {
				\draw[line width=1pt] (#2-0.1,#3-0.1) to (#2+0.1,#3+0.1);
				\node[anchor=#1] at (#2,#3) {\scriptsize 6};
			}
			\begin{tikzpicture}[yscale=-1.2]
			
			\begin{scope}[shape=circle, line width=1pt]
			\node[draw] (mult1) at (0,0) {\large$\times$};
			\node[draw] (mult2) at (3,0) {\large$\times$};
			\node[draw] (plus1) at (-0.8,1.7) {\large$+$};
			\node[draw] (plus2) at (2.2,1.7) {\large$+$};
			\end{scope}
			
			\node[draw,line width=1pt, minimum width=1.5cm, minimum height=1.5cm] (ram) at (5.8,0.7) {RAM};
			
			\node[anchor=west] (mc1) at (-2.1,-0.6) {$-\dfrac{a_{01}}{a_{00}}$};
			\node[anchor=west] (pc1) at (-2.1,0.53) {$\dfrac{b_{0}}{a_{00}}$};
			\node[anchor=west] (mc2) at (0.9,-0.6) {$-\dfrac{a_{10}}{a_{11}}$};
			\node[anchor=west] (pc2) at (0.9,0.53) {$\dfrac{b_{1}}{a_{11}}$};
			
			\begin{scope}[line width=0.6pt]
			\draw[-latex] (mult1.south) -- +(0,0.2) -| (plus1.north east);
			\draw[-latex] (mult2.south) -- +(0,0.2) -| (plus2.north east);
			
			\draw[-latex] (mc1.east) -| (mult1.north west);
			\draw[-latex] (pc1.east) -| (plus1.north west);
			\draw[-latex] (mc2.east) -| (mult2.north west);
			\draw[-latex] (pc2.east) -| (plus2.north west);
			
			\draw[-latex] (plus2.south) -- +(0,0.3) -| ([xshift=-2mm]ram.south);
			\draw[-latex] (plus1.south) -- +(0,0.4) -| ([xshift=2mm]ram.south);
			
			\draw[-latex] ([xshift=-2mm]ram.north) -- +(0,-0.7) -| (mult2.north east);
			\draw[-latex] ([xshift=2mm]ram.north) -- +(0,-1.2) -| (mult1.north east);
			\end{scope}
			
			\busmark{6.0}{1.8}
			\busmark{5.6}{1.8}
			\busmark{6.0}{-0.3}
			\busmark{5.6}{-0.3}
			\busmark[south]{2.35}{-0.6}
			\busmark[south]{-0.65}{-0.6}
			\busmarkp{-1.08}{0.95}
			\busmarkp{-0.52}{0.95}
			\busmarkp{1.92}{0.95}
			\busmarkp{2.48}{0.95}
			
			\node at (4.5, -0.85) {$x_0{\left(k, i\right)}$};
			\node at (3.4, -1.35) {$x_1{\left(k, i\right)}$};
			\node at (0.6, 2.2) {$x_0{\left(k', i'\right)}$};
			\node at (4.2, 2.1) {$x_1{\left(k', i'\right)}$};
			
			\end{tikzpicture}
					\caption{Jacobi method ($\delta = 3$).}
					\label{fig:jacobi_datapath}
				\end{subfigure}
				\hfill
				\vspace{10pt}
				\begin{subfigure}{\columnwidth}
					\centering
								\newcommand\busmark[3][west] {
				\draw[line width=1pt] (#2-0.1,#3-0.1) to (#2+0.1,#3+0.1);
				\node[anchor=#1] at (#2,#3) {\scriptsize 2};
			}
			\newcommand\busmarkp[3][west] {
				\draw[line width=1pt] (#2-0.1,#3-0.1) to (#2+0.1,#3+0.1);
				\node[anchor=#1] at (#2,#3) {\scriptsize 6};
			}
			\begin{tikzpicture}[yscale=-1.2]
			
			\begin{scope}[shape=circle, line width=1pt]
			\node[draw] (mult1) at (0.5,0) {\large$\times$};
			\node[draw] (div1) at (3,0) {\large$\div$};
			\node[draw] (plus2) at (1.75,1.7) {\large$+$};
			\end{scope}
			
			\node[draw,line width=1pt, minimum width=1.5cm, minimum height=1.5cm] (ram) at (6,0.7) {RAM};
			
			\node (mc1) at (-0.7,-0.5) {$-\dfrac{1}{2}$};
			\node (mc2) at (1.8,-0.5) {$-\dfrac{3}{2a}$};
			
			\begin{scope}[line width=0.6pt]
			\draw[-latex] (mult1.south) -- +(0,0.2) -| (plus2.north west);
			\draw[-latex] (div1.south) -- +(0,0.2) -| (plus2.north east);
			
			\draw[-latex] (mc1.east) -| (mult1.north west);
			\draw[-latex] (mc2.east) -| (div1.north west);
			
			\draw[-latex] (plus2.south) -- +(0,0.3) -| ([xshift=0mm]ram.south);
			
			\draw[-latex] ([xshift=0mm]ram.north) -- +(0,-1.05) -| (div1.north east);
			\draw[-latex] ([xshift=0mm]ram.north) -- +(0,-1.05) -| (mult1.north east);
			\end{scope}
			
			\busmark{6}{1.85}
			\busmark{6}{-0.45}
			\busmark[south]{2.5}{-0.5}
			\busmark[south]{-0.05}{-0.5}
			\busmarkp{1.48}{0.95}
			\busmarkp{2.02}{0.95}
			
			\node at (4.6, -0.7) {$x{\left(k, i\right)}$};
			\node at (3.9, 2.1) {$x{\left(k', i'\right)}$};
			
			\end{tikzpicture}
					\caption{Newton's method ($\delta = 4$).}
					\label{fig:newton}
				\end{subfigure}
				\caption{
					\architect{} benchmark datapaths.
					Adders, multipliers and dividers are arbitrary-precision radix-2 signed-digit online operators.
					Use of three-digit adders reduces online delay by 2 over their serial equivalents.
				}
			\end{figure}
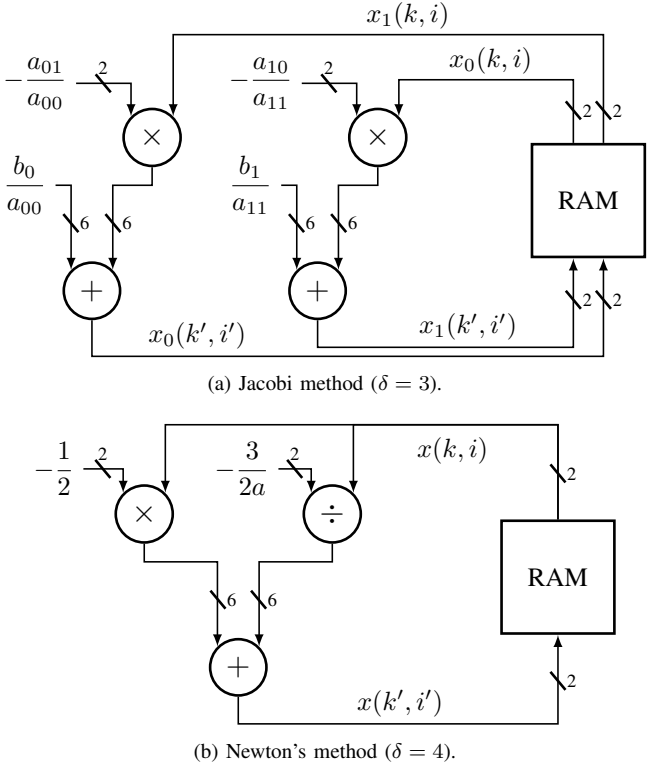

			Without loss of generality, the datapath we developed to solve systems with dimensionality $N = 2$ is depicted in Fig.~\ref{fig:jacobi_datapath}, featuring \architect{} numerical operators as described in Section~\ref{sec:architect_op}.
			Jacobi solvers with $N > 2$ could have been built with additional multipliers and adders, but this is not the emphasis---demonstrating arbitrary-accuracy iterative calculation---of this work.
			Note that runtime division is unnecessary since $\boldsymbol{A}$ and $\boldsymbol{b}$ are constants and that simple rearrangement transforms subtraction into addition.
			
		\subsection{Newton's Method}

			Newton's method is a root-finding algorithm, commonly employed to approximate the zeroes of a real-valued function $f$.
			The iterative process is
			\begin{equation*}
				x\iter{k + 1} = x\iter{k} - \frac{\func{f}{x\iter{k}}}{\func{f'}{x\iter{k}}},
			\end{equation*}
			where $f'$ is the first derivative of $f$.
			Assuming that $\func{f}{x} = 0$ is soluble and $\func{f'}{x}$ is Lipschitz continuous, convergence is quadratic if $x\iter{0}$ is sufficiently close to the solution~\cite{kelley1995iterative}.
			
			We implemented the datapath shown in Fig.~\ref{fig:newton}, again with \architect{} operators, as a second case study.
			This can solve equations of the form $\func{f}{x} = ax^2 - 3 = 0$:
			\begin{equation*}
				x\iter{k + 1} = \frac{x\iter{k}}{2} + \frac{3}{2ax\iter{k}}.
			\end{equation*}
			Since the solution of $\func{f}{x} = 0$ is irrational for some choices of $a$ (\emph{e.g.} 1), we consider this to be a particularly good showcase of \architect{}'s arbitrary-precision capabilities.
		
	\section{Evaluation}
	\label{sec:exp_results}
		
		We conducted theoretical analysis and performed experiments to investigate how \architect{} scales and performs versus competing arithmetic implementations, both traditional (LSD-first) and online, using the Jacobi and Newton's methods as benchmarks.
		Performance is evaluated in terms of latency, which, for all implementations considered in this article, is the multiplicative inverse of throughput.
		
		The closest study to this work is that presented by Zhao \emph{et al.}~\cite{fpt16}, which we compare against directly.
		For comparison against traditional arithmetic, we chose to implement parallel-in serial-out (PISO) operators since \architect{} operates in a similar digit-serial fashion.
		PISO sits at the midpoint between fully serial (SISO) and parallel (PIPO) in terms of area and performance~\cite{FPL15MUL}.
		With increase in precision $P$---which, for traditional arithmetic, can solve problems requiring precision \emph{up to} $P$---PISO suffers less from area growth and operating frequency $f_\textnormal{max}$ degradation than PIPO~\cite{meher2011high} while also being dramatically faster than SISO~\cite{serialFCCM15}.
		While we focus exclusively on hardware implementations, the limitations revealed for PISO apply equally to software libraries since precision must be chosen prior to iterative algorithmic commencement.
		
		\subsection{Complexity Analysis}
		
			In Table~\ref{tab:complexity_comparison}, we present the results of asymptotic complexity analysis---in terms of circuit size, memory requirements and latency---performed for \architect{} and its competitors.
			For PISO, we assume the repeated evaluation of an iterative expression using datapaths composed of standard numeric operators.
			For each arithmetic, we further assume latency-optimal datapath implementations featuring minimal-depth adder (for Jacobi) and multiplier (Newton) trees.
			Complexities for Zhao \emph{et al.}'s implementations were derived from analytical expressions provided by the authors~\cite{fpt16}.

			\setlength{\tabcolsep}{3.5pt}
			\begin{table}
				\caption{Complexities of iterative solver implementations.}
				\centering
				\begin{threeparttable}
						\begin{tabular}{cccc}
							\toprule
															& Area							& Memory														& Solve time																\\
							\midrule
							PISO							& $\func{\mathcal{O}}{N^2P}$	& $\func{\mathcal{O}}{NP}$, $\func{\mathcal{O}}{P}$\tnote{1}	& $\func{\mathcal{O}}{\func{\text{log}}{N}KP}$										\\
							Zhao \emph{et al.}~\cite{fpt16}	& $\func{\mathcal{O}}{N^2K}$	& $\func{\mathcal{O}}{N^2KP}$									& $\func{\mathcal{O}}{P{\left(\func{\text{log}}{N}K + P\right)}}$					\\
							\architect{}					& $\func{\mathcal{O}}{N^2}$		& $\func{\mathcal{O}}{N^2{\left(K + P\right)}^2}$				& $\func{\mathcal{O}}{\frac{\left(\func{\text{log}}{N}K + P\right)^3}{\func{\text{log}}{N}}}$	\\
							\bottomrule
						\end{tabular}
						\begin{tablenotes}
							\item[1] $N$-dimensional Jacobi method, $N^\text{th}$-order Newton's method.
						\end{tablenotes}
				\end{threeparttable}
				\label{tab:complexity_comparison}
			\end{table}
			\resettabcolsep

			Since we have chosen to analyse latency-optimised datapaths, area scales with the required number of multipliers (Newton) and adders (Jacobi), which themselves grow quadratically with $N$.
			For PISO, area also scales linearly with the width of its input operands, controlled by $P$, while the size of Zhao \emph{et al.}'s implementations instead scales linearly with the number of iterations to be performed, $K$.
			The area of an \architect{} implementation, however, scales with neither $K$ nor $P$, since the same arithmetic operators compute every approximant, to any precision, for the chosen iterative method.
			
			As with area, a PISO implementation's memory footprint scales linearly with $P$; for the Jacobi method, scaling is also linear in $N$ due to the size of the computed vector.
			Both Zhao \emph{et al.}'s implementations and \architect{} require residue storage within their multipliers and dividers; memory occupancy therefore scales with area for the arbitrary-precision architectures.
			For the former, use of memory also scales with $P$ as residues are stored to the same precision as its input data.
			Since \architect{} effectively collapses approximant and precision indices into a single dimension via its CPF, the memory requirements for each operator are determined by the maximum value of \eqref{eq:cpf} during computation to the target $\left(K, P\right)$.
			They thus scale quadratically with $K + P$.
			
			PISO's latency grows linearly with $K$ and $P$, but logarithmically with $N$ due to our aforementioned choice of adder (and multiplier) structures.
			Zhao \emph{et al.}'s speed is bottlenecked by the growth of precision---quadratically---as well as the frequency of pipeline flushes, which grows as $\func{\mathcal{O}}{\func{\text{log}}{N}KP}$~\cite{fpt16}.
			For \architect{}, given that each datapath's highest cumulative online delay $\delta$ is logarithmically related to $N$, its latency complexity can be determined by solving for $T_2$ in Section~\ref{subsubsec:comp_time_in_clk}.
			Note that $T_2$ dominates $T_1$ in all cases and $T_3 = 0$ since we assume the use of digit-parallel adders.
			
			At first glance, it appears that \architect{} behaves more poorly than its competitors in terms of memory use and solve time when scaled.
			We emphasise, however, that these complexities represent worst-case scenarios for \architect{}: optimisations including digit elision do not factor into its asymptotic behaviour but do significantly improve the average case.
			They also do not take fundamental limitations of the alternatives into account.
			In particular, exact computation to a given $\left(K, P\right)$ is rarely possible with $P$-digit LSD-first arithmetic due to rounding errors introduced in earlier approximants; only MSD-first architectures are capable of producing exact results for every approximant.
			Additionally, they do not account for \architect{}'s unique ability to compute results to \emph{any} required accuracy, effectively allowing the necessary $\left(K, P\right)$ to be determined, on a problem-by-problem basis, at runtime.
			In contrast, a PISO implementation's precision is always bounded, while the same is true of iteration count for Zhao \emph{et al.}'s proposal.
			In the remainder of this section, we empirically explore the implications of these issues.

		\subsection{Experimental Particulars}
		
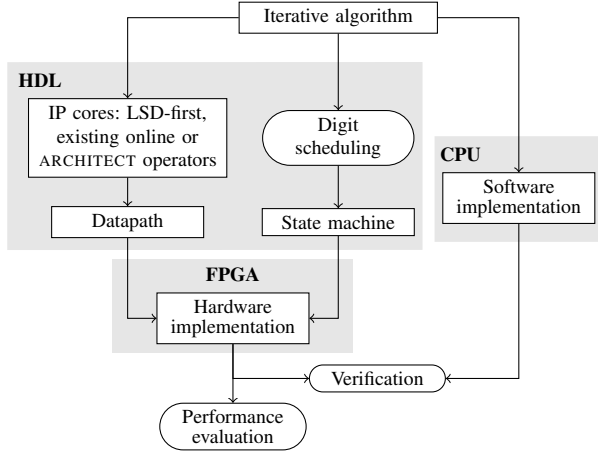
\begin{figure}
				\centering
				\newcommand\busmark[3][west] {
	\draw[line width=1pt] (#2-0.1,#3-0.1) to (#2+0.1,#3+0.1);
	\node[anchor=#1] at (#2,#3) {\scriptsize 2};
}
\newcommand\busmarkp[3][west] {
	\draw[line width=1pt] (#2-0.1,#3-0.1) to (#2+0.1,#3+0.1);
	\node[anchor=#1] at (#2,#3) {\scriptsize 6};
}

\tikzstyle{terminator} = [ draw, rounded rectangle, fill=red!20, node distance=2cm, minimum height=2em]

\begin{tikzpicture}[label distance=2mm,decoration={markings,mark= at position 0.5 with{\node[font=\footnotesize] {/};} },scale=0.8, every node/.style={scale=0.75}]

    \tikzstyle{ItemBox} = [rectangle, draw, fill=white, text centered, minimum height=1em, minimum width=10em, node distance=3cm, text width=9em]
	\tikzstyle{ItemBox2} = [rectangle, draw, fill=white, text centered, minimum height=1em, minimum width=6em, node distance=3cm, text width=7em]
    \tikzstyle{ActionCircle1} = [rounded rectangle, draw, fill=white, text centered, minimum height=1em, minimum width=9em, node distance=3cm, text width=9em]
    \tikzstyle{ActionCircle2} = [rounded rectangle, draw, fill=white, text centered, minimum height=1em, minimum width=6em, node distance=3cm, text width=6em]
    \tikzstyle{TextBox} = [rectangle, text centered, minimum height=1em, minimum width=10em, node distance=3cm, text width=9em]
    
    \node[ItemBox] at (0.0, 6.0) (IA) {Iterative algorithm};
    
    \node[ItemBox] at ($(IA)+(-3.5,-2.0)$) (IP) {IP cores: LSD-first, existing online or \architect{} operators};
    \node[ItemBox2] at ($(IA)+(3.0,-3.0)$) (SW_impl) {\shortstack{Software \\ implementation}};
    \node[ActionCircle2] at ($(IA)+(0,-2.0)$) (DS) {Digit scheduling};
        
    \node[ItemBox2] at ($(IP)+(0,-1.4)$) (datapath) {Datapath};
	\node[ItemBox2] at ($(DS)+(0,-1.4)$) (FSM) {State machine};
    
    \node[ItemBox2] at ($(IA)+(-1.75,-5.0)$) (HW_impl) {\shortstack{Hardware \\ implementation}};
    \node[ActionCircle2] at ($(IA)+(0.65,-6.0)$) (veri) {Verification};
    
    \node[ActionCircle2] at ($(HW_impl)+(0,-1.8)$) (output) {\shortstack{Performance \\evaluation} };

    \draw [->] (IA.west) -| (IP.north);
    \draw [->] (IA.south) -| (DS.north);
    \draw [->] (IP.south) -- (datapath.north);
    \draw [->] (datapath.south) |- (HW_impl.west);
    \draw [->] (DS.south) -- (FSM.north);
    \draw [->] (FSM.south) |- (HW_impl.east);
    \draw [->] (HW_impl.south) |- (veri.west);
    \draw [->] (IA.east) -| (SW_impl.north);
    \draw [->] (SW_impl.south) |- (veri.east);
    \draw [->] (HW_impl.south) -- (output.north);
    
    \node[TextBox] at ($(IP.north west)+(0.2, 0.3)$) {\bf HDL};
    \node[TextBox] at ($(HW_impl.north)+(0.0, 0.3)$) {\bf FPGA};
    \node[TextBox] at ($(SW_impl.north west)+(0.3, 0.3)$) {\bf CPU};

\begin{scope}[on background layer]
    
    \draw [draw=none, fill=black!10] ($(IP.north)+(-2.0, 0.6)$) -- ($(DS.north)+(1.4, 0.8)$) -- ($(DS.south)+(1.4, -1.4)$) -- ($(datapath.south)+(-2.0,-0.2)$) -- ($(IP.north)+(-2.0,0.5)$);
    
    \draw [draw=none, fill=black!10] ($(SW_impl.north)+(-1.4, 0.6)$) -- ($(SW_impl.north)+(1.4, 0.6)$) -- ($(SW_impl.south)+(1.4,-0.3)$) -- ($(SW_impl.south)+(-1.4,-0.3)$) -- ($(SW_impl.north)+(-1.4, 0.5)$);
    
    \draw [draw=none, fill=black!10] ($(HW_impl.north)+(-2.0, 0.55)$) -- ($(HW_impl.north)+(2.0, 0.55)$) -- ($(HW_impl.south)+(2.0,-0.15)$) -- ($(HW_impl.south)+(-2.0,-0.15)$) -- ($(HW_impl.north)+(-2.0, 0.55)$);

\end{scope}

\end{tikzpicture}
				\caption{Experimental setup for the evaluation of \architect{}.}
				\label{fig:Experiment_setup}
			\end{figure}

			We targetted a Xilinx Virtex UltraScale FPGA (XCVU190-FLGB2104-3-E) for all experiments detailed henceforward, with implementation performed using Vivado~16.4.
			The correctness of results obtained in hardware was verified via comparison against those produced by golden models executed in software.
			Fig.~\ref{fig:Experiment_setup} captures our experimental process.
			
		\subsection{Qualitative Performance Comparison}
		\label{subsec:qualitative}
		
			To evaluate performance for the Jacobi method, we considered systems in which
			\begin{equation*}
				\boldsymbol{A}_m =
				\begin{pmatrix} 
					1			&	1 - 2^{-m}	\\
					1 - 2^{-m}	&	1
				\end{pmatrix},
				\quad
				\boldsymbol{b} =
				\begin{pmatrix}
					b_0	\\
					b_1
				\end{pmatrix},
				\quad
				\boldsymbol{x}\iter{0} = \boldsymbol{0},
			\end{equation*}
			with $b_0$ and $b_1$ randomly selected from a uniform distribution in the range $\left[0, 1\right)$.
			As $m$ increases, condition number $\kappa{\left(\boldsymbol{A}_m\right)}$ also increases, indicating that higher precision $P$ will be required to generate a result of great-enough accuracy.
			We set accuracy bound $\eta = 2^{-6}$ and experimentally determined that the most ill-conditioned matrix requiring $P = 32$, a commonly encountered traditional arithmetic data width, to solve the associated system was that with $m = 25$, so we limited our experiments to $m \in \left[0, 25\right]$.
			We postulate that \architect{} should `win,' \emph{i.e.} compute the required result in less time, versus PISO either when the latter's precision $P$ is high and $\boldsymbol{A}_m$ is well conditioned or when $P$ is too low for an ill-conditioned $\boldsymbol{A}_m$ to allow convergence at all.
			For \architect{}, we used RAM size $\left(U, D\right) = \left(8, 2^{10}\right)$.
			Latencies were calculated using frequencies taken from Section~\ref{subsec:quantitative}.
			
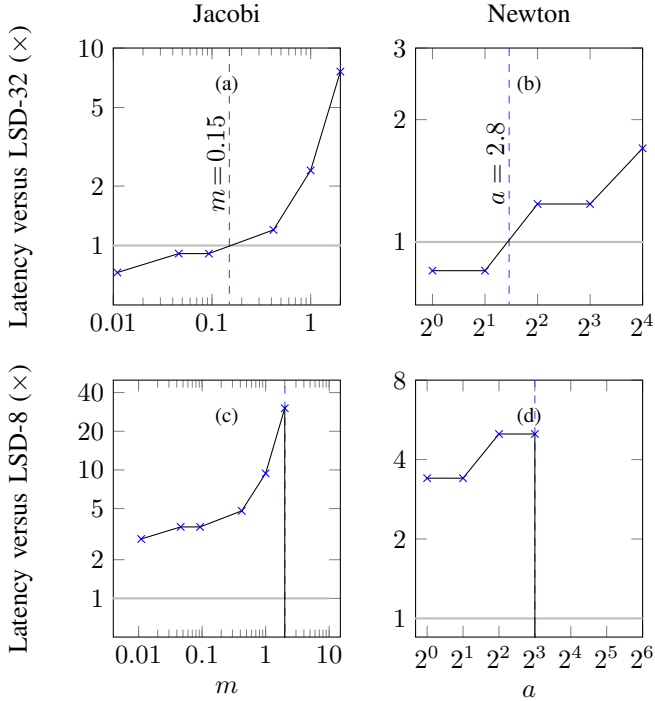
\begin{figure}
				\centering
				\begin{tikzpicture}
\newcommand\NRAB{2.8}
\newcommand\NRBC{8}
\newcommand\JAAB{0.15}
\newcommand\JABC{2.01}
\pgfplotsset{
	JABC plot/.style={mark=x, mark options={color=blue}},
	NRABC plot/.style={mark=o, mark options={color=blue}},
}
	\begin{groupplot}[
		width=46mm,
		height=50mm,
		group style={
		group size=2 by 2, xlabels at=edge bottom,
		xticklabels at=edge bottom, vertical sep=10mm},
		xlabel near ticks
	]
	
	\nextgroupplot[
		xmin=0.01,
		xmax=2,
		xtick={0.01, 0.1, 1},
		xticklabels={$0.01$,$0.1$,$1$},
		ymode=log,
		xmode=log,
        log ticks with fixed point,
		ylabel={Latency versus LSD-32 ($\times$)},
		ymin=0.5,
		ymax=10,
		ytick={1,2,5,10},
		title={Jacobi},
		]
		\addplot[thick, black!25] coordinates {(0.0001, 1) (10, 1)};
		\addplot[dashed, blue!75] coordinates {(\JAAB, 10) (\JAAB, 0.1)};
		\node[anchor=west, rotate = 90] at (axis cs:\JAAB-0.04,1.4) {$\!m\!=\! \JAAB$};		
		\addplot[JABC plot]
		table [x=m, y=TArchAB]{data/qualJA.dat};
		\node [text width=1em,anchor=north] at (axis description cs:0.5,1) {\subcaption{\label{plt:LSD_32}}};

		\nextgroupplot[
		xmin=0.8,
		xmax=16,
		xtick={1, 2, 4, 8, 16},
		xticklabels={$2^{0}$,$2^{1}$,$2^{2}$,$2^{3}$,$2^{4}$},
		ymode=log,
		xmode=log,
		log ticks with fixed point,
		ymin=0.7,
		ymax=3,
		ytick={0.5, 1, 2, 3},
		title={Newton},
		]
		\addplot[thick, black!25] coordinates {(0.0001, 1) (64, 1)};
		\addplot[dashed, blue!75] coordinates {(\NRAB-0.05, 3) (\NRAB-0.05, 0.1)};

		\node[anchor=west, rotate = 90] at (axis cs:\NRAB-0.5,1.2) {$\!a= \NRAB$};
		\addplot[JABC plot]
		table [x=aAB, y=TArchAB]{data/qualNR.dat};
		\node [text width=1em,anchor=north] at (axis description cs:0.5,1) {\subcaption{\label{plt:NRLSD_32}}};

		\nextgroupplot[
		xlabel={$m$},
		xmin=0.004,
		xmax=15,
		xtick={0.01, 0.1, 1, 10},
		ymode=log,
		xmode=log,
		log ticks with fixed point,
		ylabel={Latency versus LSD-8 ($\times$)},
		ymin=0.5,
		ymax=50,
		ytick={1,2,5,10,20, 40},
		]
		\addplot[thick, black!25] coordinates {(0.0001, 1) (10, 1)};
		\addplot[dashed, blue!75] coordinates {(\JABC, 60) (\JABC, 0.1)};
		\addplot[JABC plot]
		table [x=m, y=TArchBC]{data/qualJA.dat};
		\node [text width=1em,anchor=north] at (axis description cs:0.5,1) {\subcaption{\label{plt:LSD_8}}};

		\nextgroupplot[
		xlabel={$a$},
		xmin=0.8,
		xmax=64,
		xtick={1, 2, 4, 8, 16, 32,64},
		xticklabels={$2^{0}$,$2^{1}$,$2^{2}$,$2^{3}$,$2^{4}$,$2^{5}$,$2^{6}$},
		ymode=log,
		xmode=log,
		log ticks with fixed point,
		ymin=0.85,
		ymax=8,
		ytick={ 1, 2, 4, 8},
		]
		\addplot[thick, black!25] coordinates {(0.0001, 1) (64, 1)};
		\addplot[dashed, blue!75] coordinates {(\NRBC, 64) (\NRBC, 0.1)};
		\addplot[JABC plot]
		table [x=aBC, y=TArchBC]{data/qualNR.dat};
		\node [text width=1em,anchor=north] at (axis description cs:0.5,1) {\subcaption{\label{plt:NRLSD_8}}};

	\end{groupplot}

\end{tikzpicture}
				\caption{					
					Performance comparisons of our proposal against LSD-first arithmetic for the Jacobi and Newton's methods.
					(\subref{plt:LSD_32}) and (\subref{plt:NRLSD_32}) show how the conditioning of input matrix $\boldsymbol{A}_m$ (Jacobi) and input value $a$ (Newton) affect the solve time of our proposal compared to LSD-32.
					\architect{} computes more quickly than LSD-32 when $m \leq 0.15$ for Jacobi and $a \leq 2.8$ for Newton.
					(\subref{plt:LSD_8}) and (\subref{plt:NRLSD_8}) show that, even though our proposal leads to a slowdown compared to LSD-8, there are nevertheless points---at $m > 2$ (Jacobi) and $a > 8$ (Newton)---whence LSD-8 does not converge at all, hence our speedup is effectively infinite.
				}
				\label{plt:condition_num_precision}
			\end{figure}

			Fig.~\ref{plt:LSD_32} captures the latency ratio between \architect{} and PISO with a fixed precision of 32 bits (LSD-32) necessary to compute results for matrices with low $m$.
			Here, PISO can be said to have over-budgeted precision; results take longer to compute than had a smaller $P$ been chosen.
			For the most well conditioned matrices ($m \leq 0.15$), \architect{} takes less time to reach the target accuracy.
			For larger $m$, however, the opposite is true: lower-indexed approximants are computed to greater precision than those of PISO, taking more time.
			Had a lower choice of $P$ been made for PISO, \architect{} would have been at a disadvantage for the more well conditioned matrices, but it would also have been able to compute the results of systems featuring ill-conditioned matrices that PISO could not.
			As shown in Fig.~\ref{plt:LSD_8}, with $P = 8$ (LSD-8), \architect{} can solve systems with $m > 2$, where PISO's precision is under-budgeted; here, even if PISO ran indefinitely it would never be able to converge to an accurate-enough solution.
			We can conclude that \architect{} requires less time to generate results either when $P$ is small and convergence is fast or when $P$ is too large for PISO to ever converge.
			
			For Newton's method, we experimented with $a \in \left[1, 2^{31}\right]$.
			As $a$ increases, $\nicefrac{3}{2a}$ decreases, thus greater precision will be required for its representation.
			We calculated under termination condition $\abs{\func{f}{x\iter{k}}} < \eta$, with $\eta$ again set to $2^{-6}$.
			$a \in \left[1, 2^{31}\right]$ was chosen since, to solve $\func{f}{x}$ with $a = 2^{31}$, the worst-case precision requirement was again $P = 32$.
			
			Figs~\ref{plt:NRLSD_32} and \ref{plt:NRLSD_8} show the performance of our \architect{}-based Newton's method benchmark versus 32-bit and 8-bit PISO in the same form as Figs~\ref{plt:LSD_32} and \ref{plt:LSD_8}, respectively.
			The results achieved for Newton's method are broadly similar to those for Jacobi.
			\architect{} requires $a \leq 2.8$ to beat LSD-32 in terms of compute time, while only our proposed iterative solver can solve systems with $a > 8$ when PISO has $P = 8$.
			Identical conclusions regarding under- and over-budgeted precisions can therefore be drawn for Newton's method.
			
		\subsection{Area \& Frequency Scalability}
		\label{subsec:exp_eval}
			
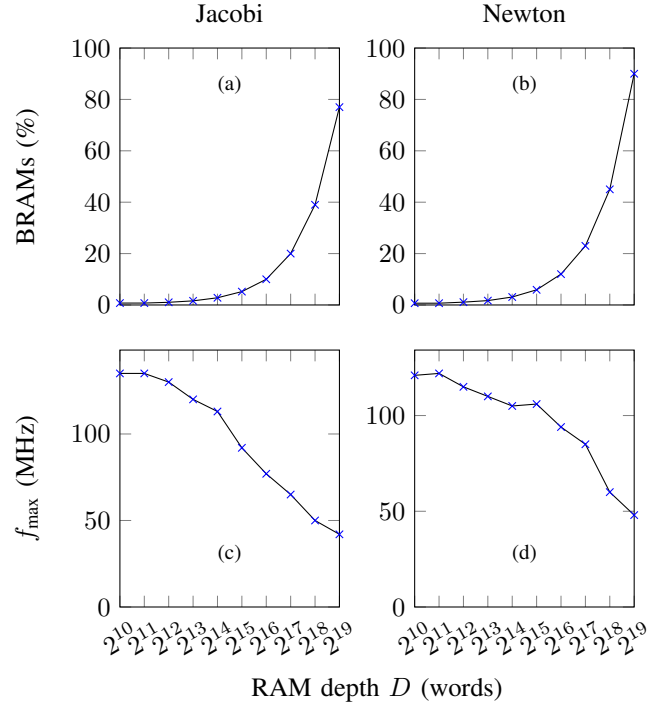
\begin{figure}
				\centering
				\makeatletter
\pgfplotsset{
	groupplot xlabel/.initial={},
	every groupplot x label/.style={
		at={($({group c1r\pgfplots@group@rows.west}|-{group c1r\pgfplots@group@rows.outer south})!0.5!({group c\pgfplots@group@columns r\pgfplots@group@rows.east}|-{group c\pgfplots@group@columns r\pgfplots@group@rows.outer south})$)},
		anchor=north,
	},
	groupplot ylabel/.initial={},
	every groupplot y label/.style={
		rotate=90,
		at={($({group c1r1.north}-|{group c1r1.outer
				west})!0.5!({group c1r\pgfplots@group@rows.south}-|{group c1r\pgfplots@group@rows.outer west})$)},
		anchor=south
	},
	execute at end groupplot/.code={%
		\node [/pgfplots/every groupplot x label]
		{\pgfkeysvalueof{/pgfplots/groupplot xlabel}};  
		\node [/pgfplots/every groupplot y label] 
		{\pgfkeysvalueof{/pgfplots/groupplot ylabel}};  
	},
	group/only outer labels/.style =
	{
		group/every plot/.code = {%
			\ifnum\pgfplots@group@current@row=\pgfplots@group@rows\else%
			\pgfkeys{xticklabels = {}, xlabel = {}}\fi%
			\ifnum\pgfplots@group@current@column=1\else%
			\pgfkeys{yticklabels = {}, ylabel = {}}\fi%
		}
	}
}

\def\endpgfplots@environment@groupplot{%
	\endpgfplots@environment@opt%
	\pgfkeys{/pgfplots/execute at end groupplot}%
	\endgroup%
}
\makeatother

\begin{tikzpicture}

	\begin{groupplot}[
		width=45mm,
		height=50mm,
		group style={
		group size=2 by 2, xlabels at=edge bottom,
		xticklabels at=edge bottom, vertical sep=6mm},
		groupplot xlabel={RAM depth $D$ (words)},
		xmode=log,
		xmin=1024,
		xmax=524288,
		xtick={1024,2048,4096,8192,16384,32768,65536,131072,262144,524288},
		xticklabels={$2^{10}$, $2^{11}$, $2^{12}$, $2^{13}$, $2^{14}$, $2^{15}$, $2^{16}$, $2^{17}$, $2^{18}$, $2^{19}$},
		xticklabel style={rotate=30},
		xlabel near ticks
	]

	\nextgroupplot[
		 ylabel={BRAMs (\%)},
        ymin=0,
        ymax = 100,
        ytick={0, 20, 40, 60, 80, 100},
        yticklabels={$0$,$20$, $40$, $60$, $80$, $100$},
        title={Jacobi},
		]
		
		
		\addplot[piso plot]
		table [x=D, y expr={(\thisrow{BRAMJA}}]{data/scalability.dat};
		\label{plt:cba_cba}
		\node [text width=1em,anchor=north] at (axis description cs:0.5,1) {\subcaption{\label{plt:bram_JA}}};

		\nextgroupplot[
		ymin=0,
		ymax = 100,
        ytick={0, 20, 40, 60, 80, 100},
		yticklabels={$0$,$20$, $40$, $60$, $80$, $100$},
		title={Newton},
		]
		
		\addplot[piso plot]
		table [x=D, y expr={(\thisrow{BRAMNR}}]{data/scalability.dat};
		\label{plt:abc_abc}
		
		\node [text width=1em,anchor=north] at (axis description cs:0.5,1) { \subcaption{\label{plt:bram_NR}}};

	\nextgroupplot[
		ylabel={$f_\textnormal{max}$ (MHz)},
		ymin=0,
		]
		
		\addplot[piso plot]
		table [x=D, y expr={\thisrow{fJA}}]{data/scalability.dat};
		\node [text width=1em,anchor=north] at (axis description cs:0.5,0.35) {\subcaption{\label{plt:fmax_JA}}};
	
			\nextgroupplot[
		ymin=0,
		]
		
		\addplot[piso plot]
		table [x=D, y expr={\thisrow{fNR}}]{data/scalability.dat};
		\node [text width=1em,anchor=north] at (axis description cs:0.5,0.35) {\subcaption{\label{plt:fmax_NR}}};

%
%
%
		
	\end{groupplot}

\end{tikzpicture}
				\caption{
					Resource use and maximum clock rate of \architect{} Jacobi and Newton benchmarks versus RAM depth $D$. 
					Area is reported in terms of BRAMs only; LUT and FF use were below 1\% for all design points.
				}
				\label{fig:scalability}
			\end{figure}

			Implementational results are presented in Fig.~\ref{fig:scalability} for our Jacobi and Newton's method benchmarks, including area and maximum operating frequency $f_\textnormal{max}$. 
			Each of the four plots features $D$, the RAM depth used for storage of each digit vector, on the $x$-axis, and RAM width $U$ was 8 in all cases.
			Lookup table (LUT) and flip-flop (FF) use are not shown since the numbers are insignificant compared to those of on-chip block RAM (BRAM)---from 0.17\% to 0.66\% for LUTs and 0.045\% to 0.21\% for FFs for the smallest ($D = 2^{10}$) and largest ($D = 2^{19}$) Jacobi designs implemented, and from 0.22\% to 0.86\% (LUTs) and 0.040\% to 0.23\% (FFs) for the Newton datapath.
			Memory use grows with $D$, as expected; the higher $K_\textnormal{res}$ and $P_\textnormal{res}$ one wishes to be able to reach, the more RAM must be instantiated.
			With 90\% and 77\% of BRAMs allocated for the Jacobi and Newton methods, respectively, we can reach $K_\textnormal{max} = 1023$ and $P_\textnormal{max} = 8184$: the maxima for our targetted FPGA with power-of-two choices of $D$.
			The small increases in non-memory resources noted can be attributed to the additional control logic and multiplexing required to address larger memories.
			The $f_\textnormal{max}$ plots show that our implementations are able to run at between 120~MHz, for the smallest $D$ tested, to around 50~MHz for the largest of both benchmarks.
			Subtle increases are due to compilation noise.

			\setlength{\tabcolsep}{4pt}
			\begin{table}
				\caption{Area-speed tradeoff via selection of RAM width $U$.}
				\centering
				\begin{tabular}{ccSSSSc}
					\toprule
					&	\multirow{2}{*}{$U$}	&	{\multirow{2}{*}{\shortstack{LUTs}}}	&	{\multirow{2}{*}{\shortstack{FFs}}}	&	{\multirow{2}{*}{\shortstack{BRAMs}}}	&	{\multirow{2}{*}{\shortstack{$f_\textnormal{max}$\\(MHz)}}}	&	\multirow{2}{*}{\shortstack{Accumulation\\latency (cycles)}}	\\
					\\
					\midrule
					\raisebox{-2pt}{\multirow{2}{*}{\rotatebox{90}{Jacobi}}}	&	8						&	1827										&	964										&	28										&	121															&	$\ceil*{\nicefrac{p\iter{k}}{8}}$						\\
					&	64						&	6964										&	2551										&	88											&	93															&	$\ceil*{\nicefrac{p\iter{k}}{64}}$						\\
					\midrule
					\raisebox{-2pt}{\multirow{2}{*}{\rotatebox{90}{Newton}}}	&	8						&	2316										&	866										&	26										&	120															&	$2\ceil*{\nicefrac{p\iter{k}}{8}} - 1$					\\
					&	64						&	6102										&	1710										&	83										&	95														&	$2\ceil*{\nicefrac{p\iter{k}}{64}} - 1$					\\
					\bottomrule
				\end{tabular}
				\label{tab:U}
			\end{table}
			\resettabcolsep
			
			\architect{} gives its users the freedom to trade off area and computation time directly by varying RAM width $U$.
			When $U$ is changed, so are the widths of the parallel online adders used in the datapath.
			While a design with narrower adders is just as able to compute a particular result as one capable of performing wider additions, it will also consume more clock cycles in return for demanding lower resource use.
			Comparisons between $U = 8$ and $U = 64$ with the same $D$, in this case $2^{10}$, are shown in Table~\ref{tab:U} to exemplify this for both of our benchmarks.
			Note that the accumulation latency for Newton's method is higher than Jacobi's due to the former's use of division; as was explained in Section~\ref{subsec:architect_div}, division requires more cycles to produce each output digit than are needed for multiplication.
			Table~\ref{tab:arith_component} shows the area breakdown and minimum clock period (critical path delay) for each of our individual arithmetic components for an example $\left(U, D\right)$.

			\begin{table}
				\caption{\architect{} operator features with RAM size $\left(U, D\right) = \left(8, 2^{10}\right)$.}
				\centering
					\begin{tabular}{cSSSS}
						\toprule
													& {\multirow{2}{*}{\shortstack{LUTs}}}	& {\multirow{2}{*}{\shortstack{FFs}}}	& {\multirow{2}{*}{\shortstack{BRAMs}}}	& {\multirow{2}{*}{\shortstack{Minimum clock\\period (ns)}}}	\\
						\\
						\midrule
						$+$							& 4										& 3										& {--}									& 2.0															\\
						$\times$					& 250									& 141									& 4										& 5.0															\\
						$\div$						& 255									& 93									& 6 									& 5.6															\\
						\bottomrule
					\end{tabular}
				\label{tab:arith_component}
			\end{table}
		
		\subsection{Quantitative Area \& Frequency Comparison}
		\label{subsec:quantitative}
			
			In order to compare the resource use and $f_\textnormal{max}$ of \architect{} against its competitors, we now assume that we wish to compute to particular $\left(K, P\right)$ targets.
			We emphasise that, since \architect{} iterates exactly while LSD-first arithmetic-based solvers do not, latency cannot be fairly compared when considering computation to a particular $\left(K, P\right)$.
			
			We chose to set targets of $\left(100, 2^{11}\right)$ (for the Jacobi method) and $\left(10, 2^{11}\right)$ (Newton).
			Thus, at their $\textnormal{100}^\textnormal{th}$ and $\textnormal{10}^\textnormal{th}$ iterations, respectively, we wish to obtain a result with 2048-digit precision.
			Fewer iterations were targetted for Newton's method due to its quadratic convergence.
			Using $U = 8$, for \architect{}, the resultant iteration counts and precisions for the two methods are $\left(K_\textnormal{res}, P_\textnormal{res}\right) = \left(509, 2545\right)$ (Jacobi) and $\left(351, 2106\right)$ (Newton).
			To successfully perform computation to $\left(K, P\right)$, we must ensure that $K_\textnormal{max} \geq K_\textnormal{res}$ and $P_\textnormal{max} \geq P_\textnormal{res}$.
			We can determine that, by setting RAM depth $D = 2^{17}$, we are able to reach	$K_\textnormal{max} = 512$ and $P_\textnormal{max} = 4088$, which satisfies these requirements for both benchmarks.

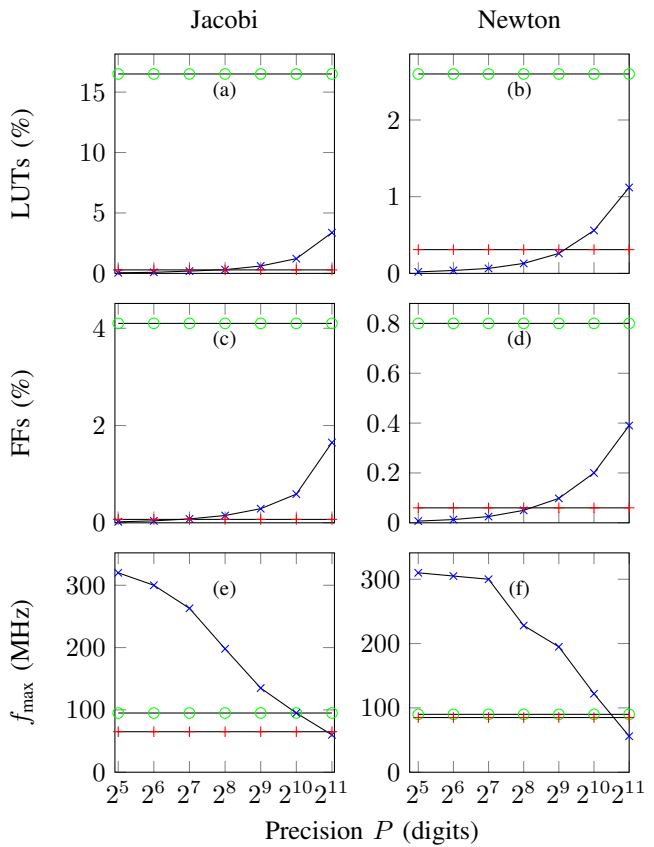
\begin{figure}
				\centering
				\makeatletter
\pgfplotsset{
	groupplot xlabel/.initial={},
	every groupplot x label/.style={
		at={($({group c1r\pgfplots@group@rows.west}|-{group c1r\pgfplots@group@rows.outer south})!0.5!({group c\pgfplots@group@columns r\pgfplots@group@rows.east}|-{group c\pgfplots@group@columns r\pgfplots@group@rows.outer south})$)},
		anchor=north,
	},
	groupplot ylabel/.initial={},
	every groupplot y label/.style={
		rotate=90,
		at={($({group c1r1.north}-|{group c1r1.outer
				west})!0.5!({group c1r\pgfplots@group@rows.south}-|{group c1r\pgfplots@group@rows.outer west})$)},
		anchor=south
	},
	execute at end groupplot/.code={%
		\node [/pgfplots/every groupplot x label]
		{\pgfkeysvalueof{/pgfplots/groupplot xlabel}};  
		\node [/pgfplots/every groupplot y label] 
		{\pgfkeysvalueof{/pgfplots/groupplot ylabel}};  
	},
	group/only outer labels/.style =
	{
		group/every plot/.code = {%
			\ifnum\pgfplots@group@current@row=\pgfplots@group@rows\else%
			\pgfkeys{xticklabels = {}, xlabel = {}}\fi%
			\ifnum\pgfplots@group@current@column=1\else%
			\pgfkeys{yticklabels = {}, ylabel = {}}\fi%
		}
	}
}

\def\endpgfplots@environment@groupplot{%
	\endpgfplots@environment@opt%
	\pgfkeys{/pgfplots/execute at end groupplot}%
	\endgroup%
}
\makeatother
\begin{tikzpicture}
	
	\begin{groupplot}[
		width=45mm,
		height=45mm,
		group style={
		group size=2 by 3, xlabels at=edge bottom,
		xticklabels at=edge bottom, vertical sep=4mm},
		groupplot xlabel={Precision $P$ (digits)},
		xtick={32,64,128,256,512,1024,2048},
		xticklabels={$2^{5}$, $2^{6}$, $2^{7}$, $2^{8}$, $2^{9}$, $2^{10}$, $2^{11}$},
		xlabel near ticks
	]
	
		\nextgroupplot[
		ylabel={LUTs (\%)},
		xmode=log,
		ymin=0,
		xmin=30,
		xmax=2150,
		title={Jacobi},
		]
		
		\addplot[fpt16 plot]
		table [x=P, y expr={(\thisrow{LUT16})}]{data/ThreeComparisonJA.dat};
		
		\addplot[piso plot]
		table [x=P, y expr={(\thisrow{LUTPISO})}]{data/ThreeComparisonJA.dat};
		
		\addplot[ours plot]
		table [x=P, y expr={(\thisrow{LUTAch})}]{data/ThreeComparisonJA.dat};
		
		\node [text width=1em,anchor=north] at (axis description cs:0.5,1) {\subcaption{\label{plt:lutJA}}};
	
		\nextgroupplot[
		xmode=log,
		ymin=0,
		xmin=27,
		xmax=2050,
		title={Newton},
		]
		
		\addplot[fpt16 plot]
		table [x=P, y expr={(\thisrow{LUT16})}]{data/ThreeComparisonNR.dat};
		\label{plt:fpt16}
		
		\addplot[piso plot]
		table [x=P, y expr={(\thisrow{LUTPISO})}]{data/ThreeComparisonNR.dat};
		\label{plt:piso}
		
		\addplot[ours plot]
		table [x=P, y expr={(\thisrow{LUTAch})}]{data/ThreeComparisonNR.dat};
		\label{plt:ours}
		\node [text width=1em,anchor=north] at (axis description cs:0.5,1) {\subcaption{\label{plt:lutNR}}};

		\nextgroupplot[
		ylabel={FFs (\%)},
		xmode=log,
		ymin=0,
		xmin=30,
		xmax=2150
		]
		
		\addplot[fpt16 plot]
		table [x=P, y expr={(\thisrow{FF16})}]{data/ThreeComparisonJA.dat};
		
		\addplot[piso plot]
		table [x=P, y expr={(\thisrow{FFPISO})}]{data/ThreeComparisonJA.dat};
		
		\addplot[ours plot]
		table [x=P, y expr={(\thisrow{FFAch})}]{data/ThreeComparisonJA.dat};
		
		\node [text width=1em,anchor=north] at (axis description cs:0.5,1) {\subcaption{\label{plt:FFJA}}};
		
			\nextgroupplot[
		xmode=log,
		ymin=0,
		xmin=27,
		xmax=2050
		]
		
		\addplot[fpt16 plot]
		table [x=P, y expr={(\thisrow{FF16})}]{data/ThreeComparisonNR.dat};
		
		\addplot[piso plot]
		table [x=P, y expr={(\thisrow{FFPISO})}]{data/ThreeComparisonNR.dat};
		
		\addplot[ours plot]
		table [x=P, y expr={(\thisrow{FFAch})}]{data/ThreeComparisonNR.dat};
		
		\node [text width=1em,anchor=north] at (axis description cs:0.5,1) {\subcaption{\label{plt:FFNR}}};

		\nextgroupplot[
		ylabel={$f_\textnormal{max}$ (MHz)},
		xmode=log,
		ymin=0,
		xmin=30,
		xmax=2150,
		]
		
		\addplot[fpt16 plot]
		table [x=P, y expr={(\thisrow{f16})}]{data/ThreeComparisonJA.dat};
		
		\addplot[piso plot]
		table [x=P, y expr={(\thisrow{fPISO})}]{data/ThreeComparisonJA.dat};
		
		\addplot[ours plot]
		table [x=P, y expr={(\thisrow{fAch})}]{data/ThreeComparisonJA.dat};
		
		\node [text width=1em,anchor=north] at (axis description cs:0.5,1) {\subcaption{\label{plt:fmax_compJA}}};
			
		\nextgroupplot[
		xmode=log,
		ymin=0,
		xmin=27,
		xmax=2050,
		]
		
		\addplot[fpt16 plot]
		table [x=P, y expr={(\thisrow{f16})}]{data/ThreeComparisonNR.dat};
		
		\addplot[piso plot]
		table [x=P, y expr={(\thisrow{fPISO})}]{data/ThreeComparisonNR.dat};
		
		\addplot[ours plot]
		table [x=P, y expr={(\thisrow{fAch})}]{data/ThreeComparisonNR.dat};
		
		\node [text width=1em,anchor=north] at (axis description cs:0.5,1) {\subcaption{\label{plt:fmax_compNR}}};	
			
	\end{groupplot}

\end{tikzpicture}
				\caption{Resource use and performance comparison of Jacobi and Newton's method implementations using Zhao \emph{et al.}'s (\ref{plt:fpt16}), PISO (\ref{plt:piso}) and our (\ref{plt:ours}) approaches versus required result precision $P$.}
				\label{fig:ThreeComparison}
			\end{figure}

			Fig.~\ref{fig:ThreeComparison} presents a side-by-side comparison of the architectures implemented following the principles presented herein and those using PISO operators as well as the online implementation published by Zhao \emph{et al.}~\cite{fpt16}.
			Most strikingly, the latter demonstrates area inefficiency, with resource use scaling linearly with iteration count $K$; \architect{} consumes 57$\times$ fewer LUTs and 59$\times$ fewer FFs than Zhao \emph{et al.}'s proposal requires in order to execute 100 iterations of the Jacobi method.
			When executing 10 iterations of Newton's method, these factors are 8.4 and 13, respectively.
			$f_\textnormal{max}$ is comparable between the two since the underlying arithmetic is largely equivalent, although \architect{}'s is slightly inferior principally due to reductions caused by the introduction of don't-change digit elision logic.
			For PISO, we can see that, while its $f_\textnormal{max}$ is initially much higher---over 300~MHz for $P = 2^4$---than \architect{}'s, it falls as $P$ increases; the crossover occurs at $P \approx 1400$.
			Taking Newton's method as an example, with a high precision requirement, such as $2^{11}$ digits, \architect{} is able to outperform its PISO counterpart in terms of $f_\textnormal{max}$ by a factor of 1.5.
			Corresponding decreases in LUT and FF use were also found: when computing to $P = 2^{10}$, again for Newton, \architect{} consumes 1.8$\times$ and 3.3$\times$ fewer of each than PISO, while for $2^{11}$ these factors increase to 3.6 and 6.5.
			Similar conclusions can be made for our implementation of the Jacobi method.
			Since the proposed designs are able to calculate to any $K \leq K_\textnormal{max}$ and $P \leq P_\textnormal{max}$, their area and $f_\textnormal{max}$ are constant.
				
		\subsection{Performance Improvement Breakdown}
		\label{subsec:performance_composite}		      
			
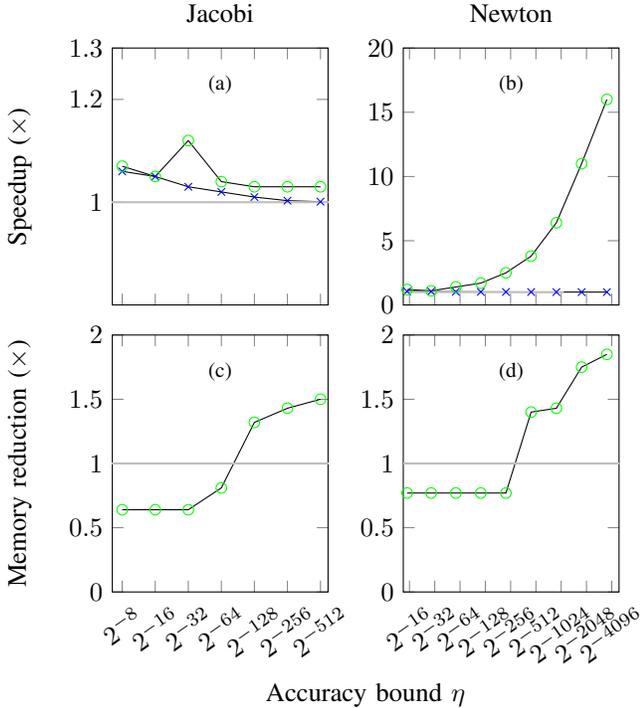
\begin{figure}
				\centering
				\makeatletter
\pgfplotsset{
	groupplot xlabel/.initial={},
	every groupplot x label/.style={
		at={($({group c1r\pgfplots@group@rows.west}|-{group c1r\pgfplots@group@rows.outer south})!0.5!({group c\pgfplots@group@columns r\pgfplots@group@rows.east}|-{group c\pgfplots@group@columns r\pgfplots@group@rows.outer south})$)},
		anchor=north,
	},
	groupplot ylabel/.initial={},
	every groupplot y label/.style={
		rotate=90,
		at={($({group c1r1.north}-|{group c1r1.outer
				west})!0.5!({group c1r\pgfplots@group@rows.south}-|{group c1r\pgfplots@group@rows.outer west})$)},
		anchor=south
	},
	execute at end groupplot/.code={%
		\node [/pgfplots/every groupplot x label]
		{\pgfkeysvalueof{/pgfplots/groupplot xlabel}};  
		\node [/pgfplots/every groupplot y label] 
		{\pgfkeysvalueof{/pgfplots/groupplot ylabel}};  
	},
	group/only outer labels/.style =
	{
		group/every plot/.code = {%
			\ifnum\pgfplots@group@current@row=\pgfplots@group@rows\else%
			\pgfkeys{xticklabels = {}, xlabel = {}}\fi%
			\ifnum\pgfplots@group@current@column=1\else%
			\pgfkeys{yticklabels = {}, ylabel = {}}\fi%
		}
	}
}

\def\endpgfplots@environment@groupplot{%
	\endpgfplots@environment@opt%
	\pgfkeys{/pgfplots/execute at end groupplot}%
	\endgroup%
}
\makeatother
\begin{tikzpicture}

	\begin{groupplot}[
	width=44.5mm,
	height=50mm,
	group style={
		group size=2 by 2, xlabels at=edge bottom,
		xticklabels at=edge bottom, vertical sep=4mm},
	groupplot xlabel={Accuracy bound $\eta$},
	xmode=log,
	xticklabel style={rotate=35},
	xlabel near ticks
	]

		\nextgroupplot[
		xmin=6.5,
		xmax=600,
		xtick={8,16,32,64,128,256,512},
		ymin=0.8,
		ymax=1.3,
		ytick={0, 1, 1.2, 1.3},
		log ticks with fixed point,
		ylabel={Speedup ($\times$)},
		title={Jacobi},
		]
		\node [text width=1em,anchor=north] at (axis description cs:0.5,1) {\subcaption{\label{plt:JA_speedup}}};
		\addplot[piso plot]
		table [x=Digits, y=factor_para]{data/time_JA.dat};
		\label{plt:paraandmsd_JA_para}
		\addplot[fpt16 plot]
		table [x=Digits, y=factor_both]{data/time_JA.dat};
		\label{plt:paraandmsd_JA_new}
		\addplot[thick, black!25] coordinates {(6, 1) (1100, 1)};

		\nextgroupplot[
			xmin=13.5,
			xmax=5000,
			xtick={16,32,64,128,256,512,1024,2048,4096},
			ymin=0,
			ymax=20,
			ytick={0, 5, 10, 15, 20},
			log ticks with fixed point,
			max space between ticks=2,			
			title={Newton},
			]
			
			\node [text width=1em,anchor=north] at (axis description cs:0.5,1) {\subcaption{\label{plt:NR_speedup}}};
			\addplot[piso plot]
			table [x=Digits, y=factor_para]{data/time_NR.dat};
			\label{plt:paraandmsd_NR_para}
			\addplot[fpt16 plot]
			table [x=Digits, y=factor_both]{data/time_NR.dat};
			\label{plt:paraandmsd_NR_new}
			\addplot[thick, black!25] coordinates {(12, 1) (1100, 1)};

		\nextgroupplot[
			xmin=6.5,
			xmax=600,
			xtick={8,16,32,64,128,256,512},
			xticklabels={$2^{-8}$, $2^{-16}$, $2^{-32}$, $2^{-64}$, $2^{-128}$, $2^{-256}$, $2^{-512}$},
			ylabel={Memory reduction ($\times$)},
			ymin=0,
			ymax=2,
			ytick={0, 0.5, 1, 1.5, 2},
			log ticks with fixed point,
			]
			
			\addplot[fpt16 plot]
			table [x=Digits, y=factor]{data/RAM_JA.dat};
			\label{plt:line_JA_new}
			\addplot[thick, black!25] coordinates {(6, 1) (1100, 1)};
			\node [text width=1em,anchor=north] at (axis description cs:0.5,1) {\subcaption{\label{plt:memJA}}};
			
		\nextgroupplot[
			xmin=13.5,
			xmax=5000,
			xtick={16,32,64,128,256,512,1024,2048,4096},
			xticklabels={$2^{-16}$, $2^{-32}$, $2^{-64}$, $2^{-128}$, $2^{-256}$, $2^{-512}$, $2^{-1024}$, $2^{-2048}$, $2^{-4096}$},
			ymin=0,
			ymax=2,
			ytick={0, 0.5, 1, 1.5, 2},
			log ticks with fixed point,
			max space between ticks=2,
			]
			
			\addplot[fpt16 plot]
			table [x=Digits, y=factor]{data/RAM_NR.dat};
			\label{plt:line_NR_new}
			\addplot[thick, black!25] coordinates {(6, 1) (11000, 1)};
			\node [text width=1em,anchor=north] at (axis description cs:0.5,1) {\subcaption{\label{plt:memNR}}};

\end{groupplot}

\end{tikzpicture}
				\caption{
					Solve time speedup for (\subref{plt:JA_speedup}) the Jacobi and (\subref{plt:NR_speedup}) Newton's methods using both don't-change digit elision and parallel online addition (\ref{plt:paraandmsd_NR_new}) and parallel addition only (\ref{plt:paraandmsd_NR_para}) versus \architect{} with both optimisations disabled.
					(\subref{plt:memJA}) and (\subref{plt:memNR}) show the corresponding memory requirement reductions for Jacobi and Newton, respectively, facilitated through digit elision.
				}
				\label{fig:performance}
			\end{figure}

			We conducted further analysis to investigate how the elision of don't-change digits and use of parallel online adders individually improve the performance and memory efficiency of \architect{}.
			Overall, Figs~\ref{plt:JA_speedup} and \ref{plt:NR_speedup} show that solve time can be significantly reduced when enabling these optimisations.
			As expected, don't-change digit elision leads to the majority of our design's efficiency savings over `vanilla' \architect{} (that without digit elision or parallel addition).
			The gaps between the don't change-plus-parallel online addition and parallel addition-only lines widen as $\eta$ is reduced, indicating that consideration of don't-change digits becomes more important with higher accuracy requirements.
			The subtle jump present in Fig.~\ref{plt:JA_speedup} is due to the $\delta$-digit granularity of elision.
			With respect to using parallel online addition only, performance is improved for higher $\eta$ since it leads to clock cycle savings when switching between iterations.
			For higher-accuracy cases, this optimisation does not contribute much to solve time speedup, however.
			This makes sense since, as $\eta$ falls, more iterations are required to achieve convergence, thus more cycles are required for the production of each new digit.
			This also affords much greater opportunity for don't-change digit elision, however, hence the high overall speedups seen on the right-hand side of, in particular, Fig.~\ref{plt:NR_speedup}.
			
			The speedups we observed for Newton's method were far more significant than those for Jacobi: up to 16$\times$ for the former.
			Relatively low performance improvements were expected for the Jacobi benchmark due to the method's linear convergence.
			Far fewer don't-change digits are detected and elided during computation than for the quadratic-convergence Newton's method, hence the less-significant latency reductions seen in Fig.~\ref{plt:JA_speedup} than Fig.~\ref{plt:NR_speedup}.
			
			Figs \ref{plt:memJA} and \ref{plt:memNR} show the memory efficiency improvements afforded through the use of don't-change digit elision for both benchmarks.
			We present these as the ratio of the number of BRAM blocks that must be instantiated on our targetted FPGA for the solution of equations to particular accuracies with and without digit elision enabled.
			The jaggedness of these plots is due to the granularity of memories, for which we only used whole numbers of BRAMs.
			For lower-accuracy cases, both pairs of designs require approximately the same amount of memory, although that considering don't-change digits is slightly inferior due to the overheads involved in comparison and subsequent elision.
			However, don't-change digit elision allows our optimised Newton design to use the same amount of memory for $\eta \leq 2^{-512}$, while vanilla \architect{} starts to consume more memory when $\eta = 2^{-449}$.
			For the test cases we evaluated, we observed up-to 1.5$\times$ and 1.9$\times$ memory savings for the Jacobi and Newton's methods, respectively.
			Beyond those shown in Fig.~\ref{fig:performance}, there are particularly high-accuracy cases---$\eta \geq 2^{-874}$ for Jacobi and $\eta \geq 2^{-7169}$ for Newton---vanilla \architect{} cannot reach before it exhausts its available memory, while that with digit elision can.
			The advantages of this scheme and its efficient memory addressing therefore come to the fore with higher accuracy requirements.
			
	\section{Conclusion \& Future Work}
	\label{sec:conslusion}
		
		In this article, we proposed the first hardware architecture capable of executing iterative algorithms to produce results of arbitrary accuracy by combining increasing iteration count with precision while using constant compute resources.
		We named this technique \architect{}.
		Our proposal employs online arithmetic to generate its results MSD first and a Cantor pairing function within its digit-storage mechanism to facilitate the simultaneous growth of iteration count and precision.
		Using digit dependency analysis, we identified stable `don't-change' digits across iterations, excluding them from calculation.
		This technique holds for any iterative method implemented using online arithmetic and was realised in hardware using simple runtime detection and digit-scheduling logic.
		We also proposed the replacement of serial online adders within iterative datapaths with parallel equivalents, facilitating latency reduction and consequent improvements in throughput.
		
		We evaluated \architect{} on FPGAs using the Jacobi and Newton's methods in order to verify its accuracy and establish its scalability and efficiency.
		These benchmarks showcased the key advantage of our approach: removing the burden of having to determine and fix the precision of arithmetic operators in advance.
		By doing so, we showed that datapaths constructed from \architect{} operators are superior to their traditional arithmetic equivalents in scenarios where the latter's precision is either overly high for the problems being solved or too low for results to converge at all.
		A single \architect{} datapath is able to compute results to any accuracy, with the only limit being imposed by the size of the available RAM.
		
		Our experiments revealed 12$\times$ LUT and 24$\times$ FF reductions over 2048-bit conventional parallel-in serial-out arithmetic, along with 57$\times$ LUT and 59$\times$ FF decreases versus the state-of-the-art online arithmetic implementation, when executing 100 Jacobi iterations.
		For Newton's method run for 10 iterations, these factors were 3.6, 6.5, 8.4 and 13, respectively.
		Versus \architect{} with the proposed don't-change and parallel addition optimisations disabled, we were able to achieve up-to 16$\times$ decreases in solve time.
		
		While we prototyped our designs on FPGAs owing to the costs and lead times associated with full-custom implementation, we note that these devices are optimised for the implementation of conventional arithmetic operators.
		In particular, FPGAs' hardened carry chains suit the construction of fast LSD-first adders.
		Our proposals cannot take advantage of such structures at present.
		We are confident that, should \architect{} see application-specific integrated circuit (ASIC) implementation, however, much more competitive performance would be achievable.
		Higher-radix ($r > 2$) online arithmetic could instead (or additionally) be employed to exploit high-performance adders, including on FPGAs, which we anticipate would also lead to speedups.
		We leave the exploration of such optimised implementations to future work.
		
		Beyond this, we will extend our benchmarking to cover additional iterative algorithms, including Krylov subspace methods such as conjugate gradient descent.
		Finally, we envisage that the arbitrary-precision computation enabled by \architect{} can be combined with high-level synthesis to enable faster hardware specialisation.

	\section*{Acknowledgements}
		
		The authors are grateful for the support of the United Kingdom EPSRC (grant numbers EP/P010040/1, EP/R006865/1 and EP/K034448/1), Imagination Technologies, the Royal Academy of Engineering and the China Scholarship Council.
		
		Supporting data for this article are available online at \texttt{https://doi.org/10.5281/zenodo.3378800}.
	
	\bibliographystyle{IEEEtran}
	\bibliography{ref_TVLSI}
	
	\begin{IEEEbiography}[{\includegraphics[width=1in,height=1.25in,clip,keepaspectratio]{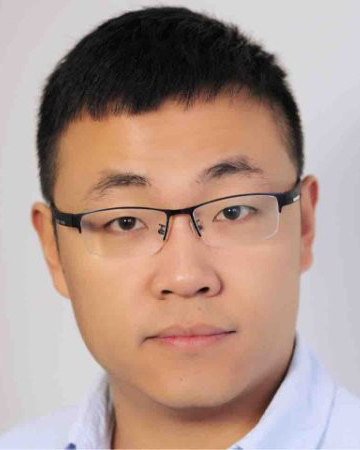}}]{He Li}
		is a PhD student in the Department of Electrical and Electronic Engineering at Imperial College London.
		He received the MS degree from the Department of Microelectronics at Tianjin University in 2016.
		His main research interests are FPGA arithmetic, custom computing and hardware security.
		He received the Best Paper Presentation Award at FPT 2017.
	\end{IEEEbiography}
	
	\begin{IEEEbiography}[{\includegraphics[width=1in,height=1.25in,clip,keepaspectratio]{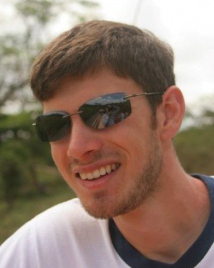}}]{James J. Davis}
		is a Research Fellow in the Department of Electrical and Electronic Engineering's Circuits and Systems group at Imperial College London.
		He received a PhD in Electrical and Electronic Engineering from Imperial College London in 2016.
		His research is focussed on the exploitation of FPGA features for cutting-edge applications, driving up performance, energy efficiency and reliability.
		Dr Davis serves on the technical programme committees of the four top-tier reconfigurable computing conferences (FPGA, FCCM, FPL and FPT) and is a multi-best paper award recipient.
		He is a Member of the IEEE and the ACM.
	\end{IEEEbiography}

	\begin{IEEEbiography}[{\includegraphics[width=1in,height=1.25in,clip,keepaspectratio]{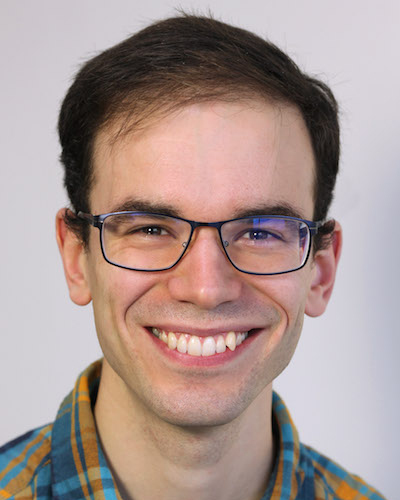}}]{John Wickerson}
		received a PhD in Computer Science from the University of Cambridge in 2013.
		He is a Lecturer in the Department of Electrical and Electronic Engineering at Imperial College London.
		His research interests include high-level synthesis, the design and implementation of programming languages and software verification.
		He is a Senior Member of the IEEE and a Member of the ACM.
	\end{IEEEbiography}

	\begin{IEEEbiography}[{\includegraphics[width=1in,height=1.25in,clip,keepaspectratio]{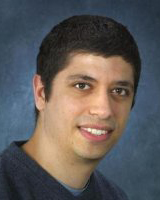}}]{George A. Constantinides}
		received the PhD degree from Imperial College London in 2001.
		Since 2002, he has been with the faculty at Imperial College London, where he is currently Professor of Digital Computation and Head of the Circuits and Systems research group.
		He was General Chair of the ACM/SIGDA International Symposium on Field-programmable Gate Arrays in 2015.
		He serves on several programme committees and has published over 200 research papers in peer-refereed journals and international conferences.
		Prof. Constantinides is a Senior Member of the IEEE and a Fellow of the British Computer Society.
	\end{IEEEbiography}

\end{document}